\documentclass[12pt]{article}

\usepackage{graphicx}
\usepackage{amsmath,amssymb}
\usepackage{bm}
\usepackage{graphicx, color}
\usepackage{wrapfig}
\begin{document}
\title{
\begin{flushright}
\ \\*[-80pt] 
\begin{minipage}{0.2\linewidth}
\normalsize
%arXiv:YYMM.NNNN \\
%KUNS-xxxx \\*[50pt]
\end{minipage}
\end{flushright}
{\Large \bf 
Tri-bimaximal Mixing and Cabibbo Angle \\in
$S_4$ Flavor Model with SUSY 
\\*[20pt]}}

\author{
\centerline{
Hajime~Ishimori$^{1,}$\footnote{E-mail address: ishimori@muse.sc.niigata-u.ac.jp},   
~Kouta~Saga$^{1,}$\footnote{E-mail address: saga@muse.sc.niigata-u.ac.jp},} \\ 
\centerline{
Yusuke~Shimizu$^{1,}$\footnote{E-mail address: shimizu@muse.sc.niigata-u.ac.jp},
~Morimitsu~Tanimoto$^{2,}$\footnote{E-mail address: tanimoto@muse.sc.niigata-u.ac.jp} }
\\*[20pt]
\centerline{
\begin{minipage}{\linewidth}
\begin{center}
$^1${\it \normalsize
Graduate~School~of~Science~and~Technology,~Niigata~University, \\ 
Niigata~950-2181,~Japan } \\
$^2${\it \normalsize
Department of Physics, Niigata University,~Niigata 950-2181, Japan } 
\end{center}
\end{minipage}}
\\*[50pt]}

\date{
\centerline{\small \bf Abstract}
\begin{minipage}{0.9\linewidth}
\medskip 
\medskip 
\small
 We present a flavor model of quarks and leptons 
with the non-Abelian discrete symmetry $S_4$ 
in the framework of the $SU(5)$ SUSY GUT.
 Three generations of $\overline 5$-plets in $SU(5)$ are assigned to ${\bf 3}$
of $S_4$ while the  first and second generations of 
$10$-plets in $SU(5)$ are assigned to ${\bf 2}$ of $S_4$, 
and the third generation of $10$-plet is assigned to ${\bf 1}$ of $S_4$.
 Right-handed neutrinos are also assigned to
 ${\bf 2}$ for the first and second generations 
and ${\bf 1}'$ for the third generation.
 We predict the Cabibbo angle 
as well as the tri-bimaximal mixing of neutrino flavors. 
We also predict the non-vanishing $U_{e3}$ of the neutrino flavor mixing 
due to higher dimensional mass operators. 
Our predicted CKM mixing angles and the $CP$ violation are 
consistent with experimental values. 
We also study SUSY breaking terms in the slepton sector. 
Our model leads to smaller values of flavor changing neutral currents 
than the present experimental bounds.
\end{minipage}
}

\begin{titlepage}
\maketitle
\thispagestyle{empty}
\end{titlepage}

\section{Introduction}

There are many free parameters in the standard model 
including its extension with neutrino mass terms 
and most of them are originated from the flavor sector, 
i.e. Yukawa couplings of quarks and leptons.
Quark masses and mixing angles have been discussed
in the standpoint of the flavor symmetries.
The discovery of neutrino masses and the  neutrino flavor mixing
 has stimulated  the work of the flavor symmetries.
Recent experiments of the neutrino oscillation 
go into a  new  phase  of precise  determination of
 mixing angles and mass squared  differences
~\cite{Schwetz:2008er,Fogli:2008jx,Fogli:2009zza,GonzalezGarcia:2010er}, 
which indicate the tri-bimaximal mixing  for three flavors 
 in the lepton sector~\cite{Harrison:2002er,Harrison:2002kp,Harrison:2003aw,Harrison:2004uh}. 
These large  mixing angles are completely  
 different from the quark mixing ones.
Therefore, it is very important
to find a natural model that leads to these mixing patterns
 of quarks and leptons with good accuracy.

%%%%%%%%%%%
The flavor symmetry is expected to explain  the mass spectrum  and
the mixing matrix of both  quarks and leptons. 
Especially, the non-Abelian discrete symmetry~\cite{Ishimori:2010au}
has been studied  intensively in the quark and lepton sectors.
 Actually, the tri-bimaximal mixing of leptons has been at first understood 
based on the non-Abelian finite group $A_4$~\cite{Ma:2001dn,Ma:2002ge,Ma:2004zv,Altarelli:2005yp,Altarelli:2005yx}.
Until now,  much progress has been made in the  theoretical
and phenomenological  analysis of $A_4$ flavor model~\cite{Babu:2002in}-\cite{delAguila:2010vg}.
On the other hand, much attention has been devoted to the question 
whether these  models  can be  extended to
describe the observed pattern of quark masses and mixing angles,
and whether these can be made compatible with the $SU(5)$ or $SO(10)$ 
grand unified theory (GUT).
The attractive candidate is the $S_4$ symmetry, which
 has been already used for the neutrino masses and the neutrino flavor mixing
~\cite{Yamanaka:1981pa,Brown:1984dk,Brown:1984mq,Ma:2005pd}.
The exact  tri-bimaximal neutrino mixing is realized
 in  $S_4$ flavor models~\cite{Lam:2008sh,Bazzocchi:2008ej,Ishimori:2008fi,Grimus:2009pg,Bazzocchi:2009pv,Bazzocchi:2009da,Meloni:2009cz}.
Many  detail studies in the  $S_4$ flavor model
have been presented  for the quark and lepton sectors~\cite{Zhang:2006fv}-\cite{Daikoku:2009pi}.
There are attempts to unify the quark and lepton sectors
 toward a grand unified theory of flavor~\cite{Hagedorn:2006ug,Cai:2006mf,Caravaglios:2005gw},
  however, quark mixing angles were not predicted clearly.

Recently, $S_4$ flavor  models to unify
 quarks and leptons have been presented 
in the framework of the $SU(5)$ SUSY GUT~\cite{Ishimori:2008fi}
or  $SO(10)$  SUSY GUT \cite{Dutta:2009bj}.
 However, quantitative  analyses have  not been given there since
the contribution from higher dimensional mass operators are not discussed. 
There also appeared the $S_4$ flavor model in $SU(5)$ SUSY GUT 
 ~\cite{Hagedorn:2010th} and the Pati-Salam SUSY GUT
\cite{Toorop:2010yh},
taking account of higher dimensional mass operators.

In this paper, we present another  $S_4$ flavor model with $Z_4$
taking account of  higher dimensional mass operators.
We predict the deviation from the tri-bimaximal mixing of
 the lepton flavor numerically.
 The CKM mixing angles and $CP$ violation are  discussed numerically
 owing  to  higher dimensional mass operators.
 We also discuss the flavor changing neutral current (FCNC) in the
SUSY sector, which is important to constrain the parameter of the 
flavor model.

The $S_4$ group has 24 distinct elements and irreducible representations 
${\bf 1},~{\bf 1}',~{\bf 2},~{\bf 3}$, and ${\bf 3}'$.
 Three generations of $\overline 5$-plets in $SU(5)$ are assigned to ${\bf 3}$
of $S_4$ while the  first and second generations of 
$10$-plets  in  $SU(5)$  are assigned to ${\bf 2}$ of $S_4$,
and the third generation of $10$-plet is  assigned to ${\bf 1}$ of $S_4$.
These  assignments of $S_4$ for $\overline 5$ and $10$ 
lead to the  completely different structure 
of  quark and lepton mass matrices.
Right-handed neutrinos, which are $SU(5)$ gauge singlets, 
are also assigned to  ${\bf 2}$ for the first and second generations,
and ${\bf 1}'$ for  the third generation.
These  assignments realize the tri-bimaximal mixing
of neutrino flavors.
Gauge singlet scalars, which are so called flavon, are introduced.
in our model.
Relevant vacuum alignment of  flavons gives the quark flavor mixing angles
  as well as the tri-bimaximal mixing of neutrino flavors.
 Especially, the Cabibbo angle is predicted to be around $15^\circ$.
%%%%%%%%%%

In section 2,
we present the $S_4\times Z_4\times U(1)_{FN}$ flavor model of
 quarks and leptons in  $SU(5)$ SUSY GUT, and 
 discuss  the  effect of the higher dimensional mass operators.  
 The deviation from the tri-bimaximal mixing is predicted,
and CKM mixing angles and the $CP$ violation are discussed in detail.
In  section 3, the alignment
of the vacuum expectation values (VEVs)  is derived.
In section 4, the FCNC in the slepton sector is discussed.
Section 5 is devoted to the summary.
We present the multiplication rule of $S_4$, 
the determination of $U(1)_{FN}$ quantum numbers, 
and the analysis of the scalar potential
in Appendices A, B, and C, respectively.

%%%%%%%%%%%%%%%%%%%%%%%%%%%%%%
%%%%%%%%%%%%%%%%%%%%%%%%%%%%%%
%%%%%%%%%%%%%%%%%%%%%%%%%%%%%%
%%%%%%%%%%%%%%%%%%%%%%%%%%%
%%%%%%%%%%%%%%%%%%%%%%%%%%%
%%%%%%%%%%%%%%%%%%%%%%%%%%%
\section{$S_4\times Z_4\times U(1)_{FN}$ flavor model with $SU(5)$ SUSY GUT}
\subsection{ Assignments of superfields}

We present the  $S_4$ flavor model in the framework of $SU(5)$ SUSY GUT. 
The flavor symmetry of quarks and leptons is the discrete group $S_4$
 in our model. The group $S_4$ has irreducible representations 
${\bf 1}$,~${\bf 1}'$,~${\bf 2}$,~${\bf 3}$, and ${\bf 3}'$. 
The multiplication rule is  shown in Appendix A.

\begin{table}[h]
\begin{tabular}{|c|ccccc||cccc|}
\hline
&$(T_1,T_2)$ & $T_3$ & $( F_1, F_2, F_3)$ & $(N_e^c,N_\mu ^c)$ & $N_\tau ^c$ & $H_5$ &$H_{\bar 5} $ & $H_{45}$ & $\Theta $ \\ \hline
$SU(5)$ & $10$ & $10$ & $\bar 5$ & $1$ & $1$ & $5$ & $\bar 5$ & $45$ & $1$ \\
$S_4$ & $\bf 2$ & $\bf 1$ & $\bf 3$ & $\bf 2$ & ${\bf 1}'$ & $\bf 1$ & $\bf 1$ & $\bf 1$ & $\bf 1$ \\
$Z_4$ & $-i$ & $-1$ & $i$ & $1$ & $1$ & $1$ & $1$ & $-1$ & $1$ \\
$U(1)_{FN}$ & $\ell $ & 0 & 0 & $m$ & 0 & 0 & 0 & 0 & $-1$ \\
\hline
\end{tabular}
\end{table}
\vspace{-0.5cm}
\begin{table}[h]
\begin{tabular}{|c|cccccc|}
\hline
& $(\chi _1,\chi _2)$ & $(\chi _3,\chi _4)$ & $(\chi _5,\chi _6,\chi _7)$ 
& $(\chi _8,\chi _9,\chi _{10})$ & $(\chi _{11},\chi _{12},\chi _{13})$ & $\chi _{14}$ \\ \hline
$SU(5)$ & $1$ & $1$ & $1$ & $1$ & $1$ & $1$ \\
$S_4$ & $\bf 2$ & $\bf 2$ & ${\bf 3}'$ & $\bf 3$ & $\bf 3$ & $\bf 1$ \\
$Z_4$ & $-i$ & $1$ & $-i$ & $-1$ & $i$ & $i$ \\
$U(1)_{FN}$ & $-\ell $ & $-n$ & 0 & 0 & 0 & $-\ell $ \\
\hline
\end{tabular}
\caption{Assignments of $SU(5)$, $S_4$, $Z_4$, and $U(1)_{FN}$ representations.}
\label{tables4}
\end{table}

Let us present the model of the quark and lepton  flavor 
with $SU(5)$ SUSY GUT. 
In $SU(5)$, matter fields are unified into $10$ 
and $\bar 5$ dimensional representations. 
Three generations of $\bar 5$, which are denoted by $F_i~(i=1,2,3)$,
 are assigned to $\bf 3$ of $S_4$. 
On the other hand, the third generation of the $10$-dimensional 
representation is assigned to $\bf 1$ of $S_4$, 
so that the top quark Yukawa coupling is allowed in the tree level. 
While, the first and second generations are assigned to $\bf 2$ of $S_4$. 
These $10$-dimensional representations are denoted by 
$T_3$ and $(T_1,T_2)$, respectively.
Right-handed neutrinos, which are $SU(5)$ gauge singlets,
are also assigned to ${\bf 1}'$ and  $\bf 2$ for $N^c_\tau$ and  
$(N^c_e,N^c_\mu)$, respectively
\footnote{Our  $S_4$ assignments of matter fields are same 
 as ones in  the model~\cite{Hagedorn:2010th}
except that right-handed neutrinos are  assigned to ${\bf 3}$ there.}.

We introduce new scalars 
$\chi_i$ in addition to the $5$-dimensional, 
$\bar 5$-dimensional,  and $45$-dimensional Higgs of $SU(5)$, $H_5$, 
$H_{\bar 5} $, and $H_{45} $, which are  assigned to $\bf 1$ of $S_4$. 
These new scalars are supposed to be $SU(5)$ gauge singlets. 
Scalars $(\chi_1,\chi_2)$ and $(\chi_3,\chi_4)$ 
are assigned to $\bf 2$, $(\chi _5,\chi _6,\chi _7)$ are assigned to ${\bf 3}'$, 
$(\chi _8,\chi _9,\chi _{10})$ and $(\chi _{11},\chi _{12},\chi _{13})$ 
are  assigned to $\bf 3$, and $\chi_{14}$ is assigned to $\bf 1$ of $S_4$
 representations, respectively. 
 In the leading order, 
 $(\chi _3,\chi _4)$ are  
coupled with the right-handed Majorana neutrino sector, 
$(\chi _5,\chi _6,\chi _7)$ are coupled with the Dirac neutrino sector, 
$(\chi _8,\chi _9,\chi _{10})$ and $(\chi _{11},\chi _{12},\chi _{13})$ 
are coupled with the charged lepton and down-type quark sectors, respectively.
 In the next-to-leading order, 
$(\chi_1,\chi_2)$ scalars 
are coupled with the  up-type  quark sector, 
and  $\chi_{14}$ contributes 
 to the charged lepton and down-type quark sectors,
 and then the mass ratio of the electron and down quark is reproduced
properly.
We also add $Z_4$ symmetry  to obtain relevant couplings.
In order to get the natural hierarchy among quark and lepton masses,
 the Froggatt-Nielsen mechanism~\cite{Froggatt:1978nt}
  is introduced as an additional 
$U(1)_{FN}$ flavor symmetry, where 
$\Theta$ denotes the Froggatt-Nielsen flavon.
The particle assignments of $SU(5)$, $S_4$, $Z_4$, and $U(1)_{FN}$
 are summarized in Table 1. 
The  $U(1)_{FN}$ charges $\ell$, $m$, and $n$
 will be determined phenomenologically.

%%%%%%%%%%%%%%%%%%%%%%%%%%%%%%%%%%%%%%%%%%%%%%%%%%%%%%%%%%%%%%%%%%%

 We can now write down  the superpotential 
respecting  $S_4$, $Z_4$, and $U(1)_{FN}$
symmetries
 in terms of the $S_4$ cutoff scale $\Lambda$,  and
the $U(1)_{FN}$ cutoff scale  $\overline \Lambda$.
The $SU(5)$ invariant superpotential 
of the Yukawa  sector up to the linear terms of $\chi_i$ ($i=1,\cdots ,13$) is given as
\begin{align}
w &= y_1^u(T_1,T_2)\otimes T_3\otimes (\chi _1,\chi _2)\otimes H_5/\Lambda + y_2^uT_3\otimes T_3\otimes H_5 \nonumber \\
&\ + y_1^N(N_e^c,N_\mu ^c)\otimes (N_e^c,N_\mu ^c)\otimes \Theta ^{2m}/\bar \Lambda ^{2m-1} \nonumber \\
&\ + y_2^N(N_e^c,N_\mu ^c)\otimes (N_e^c,N_\mu ^c)\otimes (\chi _3,\chi _4)\otimes \Theta ^{2m-n}/\bar \Lambda ^{2m-n} + MN_\tau ^c\otimes N_\tau ^c \nonumber \\
&\ + y_1^D(N_e^c,N_\mu ^c)\otimes (F_1,F_2,F_3)\otimes (\chi _5,\chi _6,\chi _7)\otimes H_5\otimes \Theta ^m/(\Lambda \bar \Lambda ^m) \nonumber \\
&\ + y_2^DN_\tau ^c\otimes (F_1,F_2,F_3)\otimes (\chi _5,\chi _6,\chi _7)\otimes H_5/\Lambda \nonumber \\
&\ + y_1(F_1,F_2,F_3)\otimes (T_1,T_2)\otimes (\chi _8,\chi _9,\chi _{10})\otimes H_{45}\otimes \Theta ^{\ell }/(\Lambda \bar \Lambda ^{\ell }) \nonumber \\
&\ + y_2(F_1,F_2,F_3)\otimes T_3\otimes (\chi _{11},\chi _{12},\chi _{13})\otimes H_{\bar 5}/\Lambda ,
\end{align}
where $y_1^u$, $y_2^u$, $y_1^N$, $y_2^N$, $y_1^D$, $y_2^D$, 
$y_1$, and $y_2$ are Yukawa couplings.
The  $U(1)_{FN}$ charges $\ell$, $m$, and $n$ are integers, and satisfy the conditions 
$m-n<0$ and $2m-n\geq 0$. 
In our numerical study, we fix  $\ell =m=1$ and $n=2$ phenomenologically
as seen in Appendix B. 
Then, some couplings are forbidden in the superpotential. 
We discuss the feature of the quark and lepton mass matrices and flavor mixing
 based on this superpotential.
However, we will take into account  the  next-to-leading 
 couplings as to  $\chi_i$  in the
 numerical study of the flavor mixing and $CP$ violation.

%%%%%%%%%%%%%%%%%%%%%%%%%%%%%%%%%%%%%%%%%%%%%%%%%%%%%%%%%%%%%%%%%%%%%%%%
%%%%%%%%%%%%%%%%%%%%%%%%%%%%%     Lepton sector    %%%%%%%%%%%%%%%%%%%%%
%%%%%%%%%%%%%%%%%%%%%%%%%%%%%%%%%%%%%%%%%%%%%%%%%%%%%%%%%%%%%%%%%%%%%%%%
\subsection{Lepton sector}

We begin to discuss the  lepton sector of 
the superpotential $w$.
 Denoting Higgs doublets as $h_u$ and $h_d$,
 the superpotential of the Yukawa sector respecting 
the $S_4 \times Z_4 \times U(1)_{FN}$ symmetry  
is given for charged leptons as
\begin{align}
w_l &=\ -3y_1\left [\frac{e^c}{\sqrt2}(l_\mu \chi _9-l_\tau \chi _{10})+\frac{\mu ^c}{\sqrt 6}(-2l_e \chi _8+l_\mu \chi _9+l_\tau \chi _{10})\right ] 
h_{45}\Theta ^{\ell }/(\Lambda \bar \Lambda ^{\ell })\nonumber \\
&\ +y_2\tau ^c (l_e \chi _{11}+l_\mu \chi _{12}+l_\tau\chi _{13})h_d/\Lambda .
\end{align}
For  right-handed Majorana neutrinos,  the superpotential is given as
\begin{align}
w_N &= y_1^N(N_e^cN_e^c+N_\mu ^cN_\mu ^c)\Theta ^{2m}/\bar \Lambda ^{2m-1}
\nonumber \\
&\ +y_2^N\left[(N_e^cN_\mu ^c+N_\mu ^cN_e^c)\chi _3+(N_e^cN_e^c-N_\mu ^cN_\mu ^c)\chi _4\right ]\Theta ^{2m-n}/\bar \Lambda ^{2m-n} + MN_\tau ^cN_\tau ^c ,
\end{align}
and for Dirac neutrino Yukawa couplings,  the superpotential is
\begin{align}
w_D &= y_1^D\left [\frac{N_e^c}{\sqrt 6}(2l_e \chi _5 -l_\mu \chi _6-l_\tau \chi _7) + \frac{N_\mu ^c}{\sqrt2}(l_\mu \chi _6-l_\tau \chi _7)\right ]
 h_u\Theta ^m/(\Lambda \bar \Lambda ^m) \nonumber \\
&\ +y_2^DN_\tau ^c(l_e\chi _5+l_\mu \chi _6+l_\tau \chi _7)h_u/\Lambda .
\end{align}
Higgs doublets $h_u$, $h_d$  and gauge singlet scalars 
$\Theta $, $\chi _i$ 
are assumed to develop their VEVs as follows:
\begin{align}
&\langle h_u\rangle =v_u,
\quad
\langle h_d\rangle =v_d,
\quad
\langle h_{45}\rangle =v_{45},
\quad
\langle \Theta \rangle =\theta , \nonumber \\
&\langle (\chi _3,\chi _4)\rangle =(u_3,u_4), 
\quad 
\langle (\chi _5,\chi _6,\chi _7)\rangle =(u_5,u_6,u_7), \nonumber \\ 
&\langle (\chi _8,\chi _9,\chi _{10})\rangle =(u_8,u_9,u_{10}),
\quad 
\langle (\chi _{11},\chi _{12},\chi _{13})\rangle =(u_{11},u_{12},u_{13}),
\label{alignment1}
\end{align}
which are supposed to be real.
Then, we obtain the mass matrix for charged leptons as
\begin{equation}
M_l = -3y_1\lambda ^\ell v_{45}\begin{pmatrix}
                       0 & \alpha _9/\sqrt 2 & -\alpha _{10}/\sqrt 2 \\
 -2\alpha _8/\sqrt 6 & \alpha _9/\sqrt 6 & \alpha _{10}/\sqrt 6   \\
                                   0 & 0 & 0 
                                \end{pmatrix}
+y_2v_d\begin{pmatrix}
               0 & 0 & 0 \\
               0 & 0 & 0 \\
               \alpha _{11} & \alpha _{12} & \alpha _{13}
          \end{pmatrix}, 
\label{charged}
\end{equation}
while the right-handed Majorana neutrino mass matrix is given as
\begin{equation}
M_N = \begin{pmatrix}
               \lambda ^{2m-n}(y_1^N\lambda ^n\bar \Lambda +y_2^N\alpha _4\Lambda ) & y_2^N\lambda ^{2m-n}\alpha _3\Lambda & 0 \\
               y_2^N\lambda ^{2m-n}\alpha _3\Lambda & \lambda ^{2m-n}(y_1^N\lambda ^n\bar \Lambda -y_2^N\alpha _4\Lambda ) & 0 \\
               0 & 0 & M
         \end{pmatrix}.
\label{majorana}
\end{equation}
Because of the condition $m-n<0$, 
 $(1,3)$, $(2,3)$, $(3,1)$, and $(3,3)$ elements of 
the right-handed Majorana neutrino mass matrix vanish. 
 These are so called SUSY zeros.
The Dirac mass matrix of neutrinos is
\begin{equation}
M_D = y_1^D\lambda ^mv_u\begin{pmatrix}
          2\alpha _5/\sqrt 6 & -\alpha _6/\sqrt 6 & -\alpha _7/\sqrt 6 \\
                            0 & \alpha _6/\sqrt 2 & -\alpha _7/\sqrt 2 \\
                            0 & 0 & 0 \end{pmatrix}+y_2^Dv_u\begin{pmatrix}
                                                               0 & 0 & 0 \\
                                                               0 & 0 & 0 \\
                                  \alpha _5 & \alpha _6 & \alpha _7
                                                             \end{pmatrix},
\label{dirac}
\end{equation}
where we denote 
$\alpha_i \equiv  u_i/\Lambda$ and $\lambda \equiv \theta /\bar \Lambda $.

In order to get the left-handed mixing of  charged leptons,
we investigate $M_l^\dagger M_l$.
If we can take   vacuum alignment 
$(u_8, u_9, u_{10})=(0, u_9, 0)$ and $(u_{11}, u_{12},  u_{13})=(0,0, u_{13})$, that is
 $\alpha _8=\alpha_{10}=\alpha_{11}=\alpha_{12}=0$,
we obtain 
\begin{equation}
M_l = \begin{pmatrix}
                                   0 & -3y_1\lambda ^\ell \alpha _9v_{45}/\sqrt 2 & 0 \\
                                   0 & -3y_1\lambda ^\ell \alpha _9v_{45}/\sqrt 6 & 0 \\
                                   0 & 0 & y_2\alpha _{13}v_d
                                \end{pmatrix},
\end{equation}
then $M_l^\dagger M_l$ is as follows:
\begin{equation}
M_l ^\dagger M_l = v_d^2
\begin{pmatrix}
0 & 0 & 0 \\
0 & 6|\bar y_1\lambda ^\ell \alpha _9|^2 & 0 \\
0 & 0 & |y_2|^2\alpha _{13}^2
\end{pmatrix},
\end{equation}
where we replace $y_1v_{45}$ with $\bar y_1v_d$. We find
 $\theta ^l_{12}=\theta^l_{13}=\theta^l_{23}=0$, where $\theta^l_{ij}$ denote
 left-handed mixing angles to diagonalize the charged lepton mass matrix.
%Since the electron mass is tiny compared with the muon mass,
% we expect  $\alpha _8 \ll \alpha_9$ and then we get the mixing angle
%$\theta_{12}^l$ as,
%\begin{align}
%\tan \theta^l_{12} &\approx -\frac{\alpha _8}{2\alpha _9},
%\end{align}
Then, charged lepton masses are 
\begin{align}
m_e^2 = 0 \ ,
\quad
m_\mu ^2 =6|\bar y_1\lambda ^\ell \alpha _9|^2v_d^2\ ,
\quad 
m_\tau ^2=|y_2|^2\alpha _{13}^2v_d^2\ .
\label{chargemass}
\end{align}
It is remarkable that the electron mass vanishes.
We will discuss the electron mass 
 as well as the down quark mass in the next-to-leading order.

%Therefore, the mixing of $\theta^l_{12}$ is estimated as
%\begin{align}
%|\tan \theta^l_{12}|\approx \frac{1}{\sqrt 3}&\frac{m_e}{m_\mu}\approx 
%2.8\times 10^{-3},
%\label{leptonmix}
%\end{align}
%which is negligibly small.
%Hereafter, we do not consider the mixing from the charged lepton sector.

 Taking vacuum alignment $(u_3, u_4)=(0, u_4)$
and   $(u_5, u_6, u_7)=(u_5, u_5, u_5)$ in Eqs.~(\ref{majorana}) and (\ref{dirac}),
the right-handed Majorana mass matrix of neutrinos turns to 
\begin{equation}
M_N = \begin{pmatrix}
               \lambda ^{2m-n}(y_1^N\lambda ^n\bar \Lambda +y_2^N\alpha _4\Lambda ) & 0 & 0 \\
               0 & \lambda ^{2m-n}(y_1^N\lambda ^n\bar \Lambda -y_2^N\alpha _4\Lambda ) & 0 \\
               0 & 0 & M
         \end{pmatrix},
\end{equation}
and the Dirac mass matrix of neutrinos becomes 
\begin{equation}
M_D = y_1^D\lambda ^mv_u\begin{pmatrix}
        2\alpha _5/\sqrt 6 & -\alpha _5/\sqrt 6 & -\alpha _5/\sqrt 6 \\
           0 & \alpha _5/\sqrt 2 & -\alpha _5/\sqrt 2 \\
                                        0 & 0 & 0 
                                     \end{pmatrix}+y_2^Dv_u\begin{pmatrix}
                                     0 & 0 & 0 \\
                                     0 & 0 & 0 \\
                         \alpha _5 & \alpha _5 & \alpha _5
                                                \end{pmatrix}.
\end{equation}

By using the seesaw mechanism $M_\nu = M_D^TM_N^{-1}M_D$, 
the left-handed Majorana neutrino mass matrix is  written as
\begin{equation}
M_\nu = \begin{pmatrix}
                 a+\frac{2}{3}b & a-\frac{1}{3}b & a-\frac{1}{3}b \\
 a-\frac{1}{3}b& a+\frac{1}{6}b+\frac{1}{2}c & a+\frac{1}{6}b-\frac{1}{2}c \\
 a-\frac{1}{3}b & a+\frac{1}{6}b-\frac{1}{2}c & a+\frac{1}{6}b+\frac{1}{2}c
            \end{pmatrix},
\label{neutrino}
\end{equation}
where
\begin{equation}
a = \frac{(y_2^D\alpha _5v_u)^2}{M},\qquad 
b = \frac{(y_1^D\alpha _5v_u\lambda ^m)^2}{\lambda ^{2m-n}(y_1^N\lambda ^n\bar \Lambda +y_2^N\alpha _4\Lambda )},\qquad 
c = \frac{(y_1^D\alpha _5v_u\lambda ^m)^2}{\lambda ^{2m-n}(y_1^N\lambda ^n\bar \Lambda -y_2^N\alpha _4\Lambda )}.
\label{neutrinomassparameter}
\end{equation}
The neutrino mass matrix is decomposed as
\begin{equation}
M_\nu = \frac{b+c}{2}\begin{pmatrix}
                        1 & 0 & 0 \\
                        0 & 1 & 0 \\
                        0 & 0 & 1
                      \end{pmatrix} + \frac{3a-b}{3}\begin{pmatrix}
                                                      1 & 1 & 1 \\
                                                      1 & 1 & 1 \\
                                                      1 & 1 & 1
             \end{pmatrix} + \frac{b-c}{2}\begin{pmatrix}
     1 & 0 & 0 \\
      0 & 0 & 1 \\
     0 & 1 & 0
     \end{pmatrix},
\end{equation}
which gives the tri-bimaximal mixing matrix 
$U_\text{tri-bi}$ and mass eigenvalues  as follows:
\begin{equation}
U_\text{tri-bi} = \begin{pmatrix}
               \frac{2}{\sqrt{6}} &  \frac{1}{\sqrt{3}} & 0 \\
     -\frac{1}{\sqrt{6}} & \frac{1}{\sqrt{3}} &  -\frac{1}{\sqrt{2}} \\
      -\frac{1}{\sqrt{6}} &  \frac{1}{\sqrt{3}} &   \frac{1}{\sqrt{2}}
         \end{pmatrix},
\qquad m_1 = b\ ,\qquad m_2 = 3a\ ,\qquad m_3 = c\ .
\label{mass123}
\end{equation}

%%%%%%%%%%%%%%%%%%%%%%%%%%%%%%%%%%%%%%%%%%%%%%%%%%%%%%%
%%%%%%%%%%%%  next-to-leading Lepton %%%%%%%%%%%%%%%%%%%%%
%%%%%%%%%%%%%%%%%%%%%%%%%%%%%%%%%%%%%%%%%%%%%%%%%%%%%%%

 The next-to-leading terms of the superpotential are important
to predict the deviation from the tri-bimaximal mixing of leptons,
especially, $U_{e3}$.  
 The relevant superpotential in the charged lepton sector
 is given at the next-to-leading order  as 
\begin{align}
\Delta w_l&=y_{\Delta _a}(T_1,T_2)\otimes (F_1,F_2,F_3)\otimes (\chi _1,\chi _2)\otimes (\chi _{11},\chi _{12},\chi _{13})\otimes H_{\bar 5}/\Lambda ^2 \nonumber \\
&\ +y_{\Delta _b}(T_1,T_2)\otimes (F_1,F_2,F_3)\otimes (\chi _5,\chi _6,\chi _7)\otimes \chi _{14}\otimes H_{\bar 5}/\Lambda ^2 \nonumber \\
&\ +y_{\Delta _c}(T_1,T_2)\otimes (F_1,F_2,F_3)\otimes (\chi _1,\chi _2)\otimes (\chi _5,\chi _6,\chi _7)\otimes H_{45}/\Lambda ^2 \nonumber \\
&\ +y_{\Delta _d}(T_1,T_2)\otimes (F_1,F_2,F_3)\otimes (\chi _{11},\chi _{12},\chi _{13})\otimes \chi _{14}\otimes H_{45}/\Lambda ^2 \nonumber \\
&\ +y_{\Delta _e}T_3\otimes (F_1,F_2,F_3)\otimes (\chi _5,\chi _6,\chi _7)\otimes (\chi _8,\chi _9,\chi _{10})\otimes H_{\bar 5}\otimes /\Lambda ^2 \nonumber \\
&\ +y_{\Delta _f}T_3\otimes (F_1,F_2,F_3)\otimes (\chi _8,\chi _9,\chi _{10})\otimes (\chi _{11},\chi _{12},\chi _{13})\otimes H_{45}\otimes /\Lambda ^2\ .
\label{nextsusy}
\end{align} 
By using this superpotential,
 we obtain the charged lepton mass matrix as follows:
\begin{equation}
M_l\simeq
\begin{pmatrix}
\epsilon _{11} & \frac{\sqrt 3m_\mu}{ 2}+\epsilon_{12} & \epsilon _{13} \\
\epsilon _{21} & \frac{m_\mu}{2}+\epsilon_{22} & \epsilon _{23} \\
\epsilon _{31} & 0 & m_\tau+\epsilon_{33}
\end{pmatrix},
\label{nextleading}
\end{equation}
where $m_\mu$ and $m_\tau$ are given in  Eq.~(\ref{chargemass}),
and $\epsilon_{ij}$'s  are calculated by using  Eq.~(\ref{nextsusy})  as
\begin{align}
\epsilon_{11}&=y_{\Delta _b}\alpha _5\alpha _{14}v_d
   -3\bar y_{\Delta _{c_2}}\alpha _1\alpha _5v_d , \nonumber \\
\epsilon_{12}&=
   -\frac{1}{2}y_{\Delta _b}\alpha _5\alpha _{14} v_d   
   +3\left [  \frac{\sqrt 3}{4}(\sqrt 3-1)\bar y_{\Delta _{c_1}}-\frac{1}{4}(\sqrt 3+1)\bar y_{\Delta _{c_2}}  
   \right ]\alpha _1\alpha _5v_d , \nonumber \\
\epsilon_{13}&=\left [\left \{ \frac{\sqrt 3}{4}(\sqrt 3-1)y_{\Delta _{a_1}}+\frac{1}{4}(\sqrt 3+1)y_{\Delta _{a_2}}\right \} \alpha _1\alpha _{13} 
   -\frac{1}{2}y_{\Delta _b}\alpha _5\alpha _{14}\right ]v_d  \nonumber \\
   &\ -3\left [\left \{ -\frac{\sqrt 3}{4}(\sqrt 3+1)\bar y_{\Delta _{c_1}}-\frac{1}{4}(\sqrt 3-1)\bar y_{\Delta _{c_2}}\right \}\alpha _1\alpha _5
   +\frac{\sqrt 3}{2}\bar y_{\Delta _d}\alpha _{13}\alpha _{14}\right ]v_d , \nonumber \\
\epsilon_{21}&=-3 \bar y_{\Delta _{c_1}}\alpha _1\alpha _5v_d , \nonumber \\
\epsilon_{22}&=\frac{\sqrt 3}{2}y_{\Delta _b}\alpha _5\alpha _{14} v_d
   +3\left [ \frac{1}{4}(\sqrt 3-1)\bar y_{\Delta _{c_1}}
   +\frac{\sqrt 3}{4}(\sqrt 3+1)\bar y_{\Delta _{c_2}}  
   \right ]\alpha _1\alpha _5 v_d , \nonumber \\
\epsilon_{23}&=\left [\left \{ -\frac{1}{4}(\sqrt 3-1)y_{\Delta _{a_1}}+\frac{\sqrt 3}{4}(\sqrt 3+1)
    y_{\Delta _{a_2}}\right \} \alpha _1\alpha _{13}
   -\frac{\sqrt 3}{2}y_{\Delta _b}\alpha _5\alpha _{14}\right ]v_d  \nonumber \\
   &\ -3\left [\left \{ \frac{1}{4}(\sqrt 3+1)\bar y_{\Delta _{c_1}}
   -\frac{\sqrt 3}{4}(\sqrt 3-1)\bar y_{\Delta _{c_2}}\right \} \alpha _1\alpha _5
   -\frac{1}{2}\bar y_{\Delta _d}\alpha _{13}\alpha _{14}\right ]v_d , \nonumber \\
\epsilon_{31}&=-y_{\Delta _e}\alpha _5\alpha _9v_d-3\bar y_{\Delta _f}\alpha _9\alpha _{13}v_d , \nonumber \\
%\epsilon_{32}&=0, \nonumber \\
\epsilon_{33}&=y_{\Delta _e}\alpha _5\alpha _9v_d .
\label{correction}
\end{align}
Magnitudes of $\epsilon_{ij}$'s are  of ${\cal O}(\tilde\alpha^2)$,
where $\tilde\alpha$ is a linear combination of $\alpha_i$'s.

Then, $M_l^\dagger M_l$  is given in terms of 
 $\epsilon_{ij}$, which give the non-vanishing
electron mass, as follows:
\begin{equation}
M_l^\dagger M_l\simeq 
\begin{pmatrix}
|\epsilon _{11}|^2 + |\epsilon _{21}|^2 + |\epsilon _{31}|^2 & \frac{1}{2}(\sqrt{3}\epsilon _{11}^*+\epsilon_{21}^*)m_\mu & \epsilon _{31}^*m_\tau \\
\frac{1}{2}(\sqrt{3}\epsilon _{11}+\epsilon _{21})m_\mu & m_\mu^2 & \frac{1}{2}(\sqrt{3}\epsilon _{13}+\epsilon _{23})m_\mu \\
\epsilon _{31}m_\tau & \frac{1}{2}(\sqrt{3}\epsilon _{13}^*+\epsilon _{23}^*)m_\mu & m_\tau^2
\end{pmatrix}.
\end{equation}
Thus, the charged lepton mass matrix is not diagonal due to 
next-to-leading terms $\epsilon_{ij}$, which give the non-vanishing
electron mass.
Since we have 
$m_\mu={\cal O}(\lambda \tilde\alpha)$, $m_\tau={\cal O}(\tilde\alpha)$, 
and $\epsilon_{ij}={\cal O}(m_e)$,
mixing angles $\theta _{12}^l$, $\theta _{13}^l$ and 
$\theta _{23}^l$ are given as 
\begin{equation}
\theta _{12}^l=\mathcal{O}\left (\frac{m_e }{m_\mu }\right ),\qquad 
\theta _{13}^l=\mathcal{O}\left (\frac{m_e }{m_\tau }\right ),\qquad 
\theta _{23}^l=\mathcal{O}\left (\frac{m_e m_\mu }{m^2_\tau }\right )\ .
\end{equation}
Therefore, the charged lepton mixing matrix is written as
\begin{equation}
U_E=
\begin{pmatrix}
1 & \mathcal{O}\left (\frac{m_e}{m_\mu}\right ) & \mathcal{O}\left (\frac{m_e}{m_\tau}\right ) \\
\mathcal{O}\left (\frac{m_e}{m_\mu}\right ) & 1 & \mathcal{O}\left (\frac{m_e m_\mu }{m^2_\tau }\right ) \\
\mathcal{O}\left (\frac{m_e}{m_\tau} \right ) & \mathcal{O}\left (\frac{m_e m_\mu }{m^2_\tau }\right ) & 1
\end{pmatrix}.
\end{equation}
Now, the lepton mixing matrix $U$ is deviated 
from the tri-bimaximal mixing as follows:
\begin{equation}
U=U_E^\dagger U_\text{tri-bi}.
\end{equation}
The lepton mixing matrix elements $U_{e3},\ U_{e2}$, and $U_{\mu 3}$ are given as
\begin{equation}
\left |U_{e3}\right |\sim \frac{1}{\sqrt 2}\left (\mathcal{O}\left (\frac{m_e}{m_\mu}\right )\right ),\ \
\left |U_{e2}\right |\sim \frac{1}{\sqrt 3}\left (1+\mathcal{O}\left (\frac{m_e}{m_\mu}\right )\right ),\ \
\left |U_{\mu 3}\right |\sim \frac{1}{\sqrt 2}\left (1-\mathcal{O}\left (\frac{m_e m_\mu }{m^2_\tau }\right )\right )\ .
\label{deviation}
\end{equation}
Thus, the deviation from the tri-bimaximal mixing is lower than 
 $\mathcal{O}\left (0.01\right )$, which is rather small.

Let us  discuss the electron and down quark masses. 
The determinant of $M_l^\dagger M_l$ is 
\begin{equation}
\det \left [M_l^\dagger M_l\right ] 
\simeq \frac32 m_\mu^2m_\tau^2\left (\frac{1}{6}\epsilon _{11}^2-\frac{1}{\sqrt 3}\epsilon _{11}\epsilon _{21}+\frac{1}{2}\epsilon _{21}^2\right )\ ,
\end{equation}
where $\epsilon_{ij}$'s are taken to be real for simplicity.
Then the electron mass is given as
\begin{align}
m_e^2 &\simeq \frac{3}{2}\left (\frac{1}{6}\epsilon _{11}^2-\frac{1}{\sqrt 3}\epsilon _{11}\epsilon _{21}+\frac{1}{2}\epsilon _{21}^2\right ) \nonumber \\
&\simeq \frac{3}{2}\left [\frac{1}{6}y_{\Delta _d}^2\alpha _5^2\alpha _{14}^2
+y_{\Delta _d}(\sqrt 3\bar y_{\Delta _{e_1}}-\bar y_{\Delta _{e_2}})\alpha _1\alpha _5^2\alpha _{14}
+\frac{1}{2}\left (3\bar y_{\Delta _{e_1}}-\sqrt 3\bar y_{\Delta _{e_2}}\right )^2\alpha _1^2\alpha _5^2\right ]v_d^2\ .
\label{emass}
\end{align}
In the same way, the down quark mass, which is discussed in subsection 2.3, is obtained as 
\begin{equation}
m_d^2\simeq \frac{3}{2}\left [\frac{1}{6}y_{\Delta _d}^2\alpha _5^2\alpha _{14}^2
-\frac{1}{3}y_{\Delta _d}(\sqrt 3\bar y_{\Delta _{e_1}}-\bar y_{\Delta _{e_2}})\alpha _1\alpha _5^2\alpha _{14}
+\frac{1}{18}\left (3\bar y_{\Delta _{e_1}}-\sqrt 3\bar y_{\Delta _{e_2}}\right )^2\alpha _1^2\alpha _5^2\right ]v_d^2\ .
\label{dmass}
\end{equation}
In order to get  the ratio  
$m_e^2:m_d^2=1:9$,
%which is consistent with the observed mass spectra at the GUT scale,
 we require the following condition:
\begin{align}
\alpha _{14}=-\frac{5\left (\sqrt 3\bar y_{\Delta _{e_1}}-\bar y_{\Delta _{e_2}}\right )}{y_{\Delta _d}}\alpha _1,\quad \text{or} \quad 
\alpha _{14}=-\frac{2\left (\sqrt 3\bar y_{\Delta _{e_1}}-\bar y_{\Delta _{e_2}}\right )}{y_{\Delta _d}}\alpha _1.
\end{align}
Thus, the flavon $\chi_{14}$ is  introduced in our model
to explain the proper ratio of the electron mass and the down quark mass
although those masses appear at the next-to-leading order.

%%%%%%%%%%%%%%%%%%%%%%%%%%%%%%%%%
%%%%%%%%%%%%%%%%%%%%%%%%%%%%%%%%%
%%%%%%%%% Neutrino %%%%%%%%%%%%%%
%%%%%%%%%%%%%%%%%%%%%%%%%%%%%%%%%
 Hereafter, we fix
$\ell=1$, $m=1$, and $n=2$ as Frogatt-Nielsen charges, which are  given in
  Appendix B. 
The superpotential of the next-to-leading order for Majorana neutrinos is 
\begin{align}
\Delta w_N &= y_{\Delta _1}^N(N_e^c,N_\mu ^c)\otimes (N_e^c,N_\mu ^c)\otimes (\chi _1,\chi _2)\otimes \chi _{14}/\Lambda \nonumber \\
&\ + y_{\Delta _2}^N(N_e^c,N_\mu ^c)\otimes N_\tau ^c\otimes (\chi _5,\chi _6,\chi _7)
\otimes (\chi _{11},\chi _{12},\chi _{13})\otimes \Theta /(\Lambda \bar \Lambda ) \nonumber \\
&\ + y_{\Delta _3}^N(N_e^c,N_\mu ^c)\otimes N_\tau ^c\otimes (\chi _8,\chi _9,\chi _{10})
\otimes (\chi _8,\chi _9,\chi _{10})\otimes \Theta /(\Lambda \bar \Lambda ) \nonumber \\
&\ + y_{\Delta _4}^NN_\tau ^c\otimes N_\tau ^c\otimes (\chi _8,\chi _9,\chi _{10}) \otimes (\chi _8,\chi _9,\chi _{10})/\Lambda .
\end{align}
The dominant matrix elements of Majorana neutrinos
at the  next-to-leading order are written as follows:
\begin{align}
&\Delta M_N=\Lambda \times \nonumber \\
&\begin{pmatrix}
y_{\Delta _1}^N\alpha _1\alpha _{14} & y_{\Delta _1}^N\alpha _1\alpha _{14} 
& -\frac{\lambda}{\sqrt 6}y_{\Delta _2}^N \alpha _5\alpha _{13}+\frac{\lambda}{\sqrt 2}y_{\Delta _3}^N\lambda \alpha _9^2 \\
y_{\Delta _1}^N\alpha _1\alpha _{14} & -y_{\Delta _1}^N\alpha _1\alpha _{14} 
& -\frac{\lambda}{\sqrt 2}y_{\Delta _2}^N \alpha _5\alpha _{13}+\frac{\lambda}{\sqrt 6}y_{\Delta _3}^N \alpha _9^2 \\
-\frac{\lambda}{\sqrt 6}y_{\Delta _2}^N \alpha _5\alpha _{13}+\frac{\lambda}{\sqrt 2}y_{\Delta _3}^N \alpha _9^2 
& -\frac{\lambda}{\sqrt 2}y_{\Delta _2}^N \alpha _5\alpha _{13}+\frac{\lambda}{\sqrt 6}y_{\Delta _3}^N \alpha _9^2 & y_{\Delta _4}^N\alpha _9^2
\end{pmatrix}.
\label{majoranacorr}
\end{align}
Then, the $U_{e3}$ is estimated as 
\begin{equation}
U_{e3}\sim \frac{y_{\Delta _1}^N\alpha _1\alpha _{14}}{y_2^N\alpha _4}
\sim \mathcal{O}(\tilde \alpha )\ .
\label{ue31}
\end{equation}

We also consider the Dirac neutrino mass matrix. 
The superpotential at the next-to-leading order for Dirac neutrinos is 
given as
\begin{equation}
\Delta w_D =y_\Delta ^D(N_e^c,N_\mu ^c)\otimes (F_1,F_2,F_3)\otimes (\chi _8,\chi _9, \chi _{10})\otimes (\chi _{11},\chi _{12}, \chi _{13})
\otimes H_5\otimes \Theta /(\Lambda ^2\bar \Lambda )\ .
\end{equation}
The dominant matrix elements of the Dirac neutrinos 
at the next-to-leading order are written as follows:
\begin{equation}
\Delta M_D=
\begin{pmatrix}
\ast & \ast & \ast \\
y_{\Delta }^D\lambda \alpha _9\alpha _{13}v_u & \ast & \ast \\
\ast & \ast & \ast 
\end{pmatrix}.
\label{diraccorr}
\end{equation}
Then,  we can estimate  $U_{e3}$  as 
\begin{equation}
U_{e3}\sim -\frac{\sqrt 6y_{\Delta }^D\alpha _9\alpha _{13}}{3y_1^D\alpha _5}\sim \mathcal{O}(\tilde \alpha )\ .
\label{ue32}
\end{equation}
Thus, the contribution of the next-to-leading terms on $U_{e3}$
is of ${\cal O}(\tilde \alpha )$ in the neutrino sector
while that is  ${\cal O}(m_e/m_\mu )$ in the charged lepton  sector.
Therefore, it is concluded that the deviation from the tri-bimaximal mixing 
mainly comes from  the neutrino sector.

%%%%%%%%%%%%%%%%%%%%%%%%%%%%%%%%%%%%%%%%%%%
%%%%%%%%%%%% Numerical Results %%%%%%%%%%%%
%%%%%%%%%%%%%%%%%%%%%%%%%%%%%%%%%%%%%%%%%%%
%%%%%%%%%%%%%%%%%%%%%%%%%%%%%%%%%%%%%%%%%%%%%%%%%%%%%%%%%%%%%%%%%%%%%%%%%%%%%%%%%
\begin{figure}[htb]
\begin{minipage}[]{0.4\linewidth} 
\includegraphics[width=7cm]{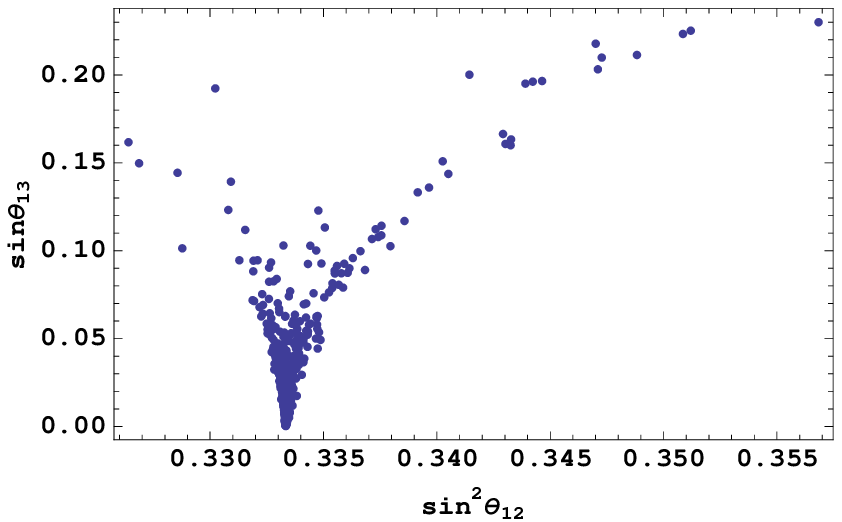}
\caption{
The allowed region on  $\sin ^2\theta _{12}$--$\sin\theta _{13}$ plane.}
\end{minipage}
\hspace{2cm}
\begin{minipage}[]{0.4\linewidth} 
\includegraphics[width=7 cm]{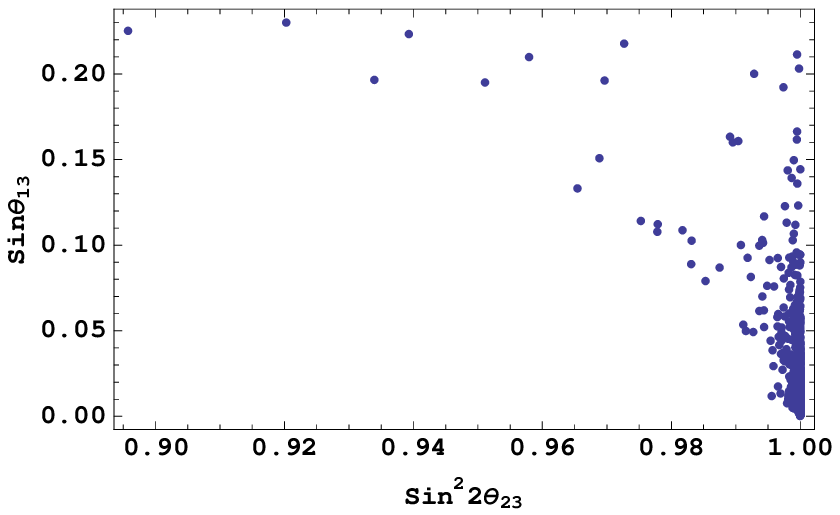}
\caption{
The allowed region on  $\sin ^22 \theta _{23}$--$\sin\theta _{13}$ plane.}
\end{minipage}
\end{figure}

%%%%%%%%%%%%%%%%%%%%%%%%%%%%%%%%%%%%%%%%%%%%%%%%%%%%%%%%%%%%%%%%%%%%%%%%%%%%
Let us discuss the deviation from the tri-bimaximal mixing numerically.
In order to obtain numerical result,
all Yukawa couplings at the leading order  are  complex,  
and these absolute values and phases  are taken to be randomly chosen 
from $0$ to $1$ and $-\pi $ to $\pi $, respectively. Other Yukawa couplings
 at the next-to-leading order are to be real and  randomly chosen 
from $-1$ to $1$ since the contribution from these phases 
on the $CP$ violation  is very small.
 Parameters $\alpha_i$'s are fixed 
as seen in section 3. We also take
$M=10^{12}$ GeV. 
The  parameter $\alpha_{14}$ is constrained to reproduce the proper ratio
of the electron mass and down quark mass as seen 
in Eqs.~(\ref{emass}) and (\ref{dmass}). Yukawa couplings
are also constrained to give  the absolute value of the electron mass.

We present  numerical  result of the deviation from the tri-bimaximal mixing
 with scattering plots. 
Here, we neglect the renomalization effect of the 
neutrino mass matrix because we suppose the normal hierarchy of 
neutrino masses and take small $\tan\beta ~(=3)$.
%%%%%%%%%%%%%%%%%%%%%%%%%%
\begin{wrapfigure}{r}{7cm}
\begin{center}
\includegraphics[width=7cm]{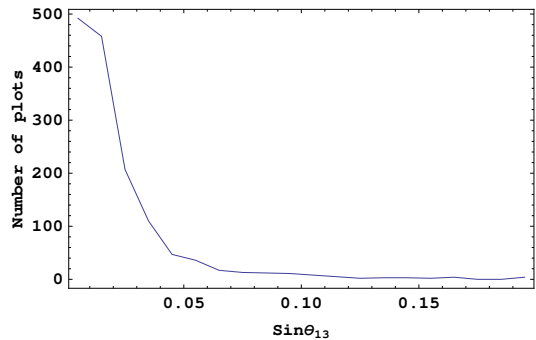}
\caption{Number of points   versus  $\sin\theta _{13}$.}
\end{center}
\end{wrapfigure}

%%%%%%%%%%%%%%%%%%%%%%%%%%%%%%%%
%%%%%%%%%%%%%%%%%%%%%%%%%%%%%%%%
We show our prediction of $\sin\theta_{13}$ 
versus  $\sin^2\theta_{12}$ in Figure 1, where $\theta_{ij}$'s
are lepton mixing angles in the usual convention. That is,
 $\sin\theta_{13}=|U_{e3}|$. 
In Figure 2, we show  the prediction of  $\sin\theta_{13}$ on the   
$\sin^2 2\theta_{23}$--$\sin^2\theta_{13}$ plane. 
The predicted upper bound of $\sin\theta_{13}$ could be larger than $0.1$
by tuning Yukawa couplings. 
 In the case that $\sin\theta_{13}$ is  larger than $0.15$, 
the value of $\sin^2\theta_{12}$ considerably deviates from  
the tri-maximum value $1/3$,
that is  $\sin^2\theta_{12}\geq 0.34$. 
However, the predicted points  of $\sin\theta_{13}$ distribute   
mainly in the region lower than $0.07$. 
The mixing angle $\theta_{23}$
 is  mainly  predicted near the maximal mixing angle  $\pi/4$.
The predicted points larger than $0.99$ for $\sin^2 2\theta_{23}$ cover $99\%$.

We investigate  precisely the predicted value 
 $\sin\theta_{13}$  in our model.
Let us calculate  the expectation value of $\theta_{13}$.
In our calculation, one million parameter sets are generated randomly.
The number of allowed parameter sets in experimental constraints is 
$1442$. These points have been plotted   in Figures 1 and 2. 
In Figure 3, we present the distribution of the plot versus  $\sin \theta_{13}$.
By using this result, we can  calculate the mean value of  $\theta_{13}$ 
and  the standard deviation. 
%The mean value and standard deviation are 
%given by $\langle x\rangle=\frac1n\sum_{i}x_i$ 
%and $\sigma^2=\frac1n\sum_i(x_i-\langle x\rangle)^2$. 
The mean value of  $\sin\theta_{13}$ is $0.023$
and the standard deviation is $0.028$. 
Thus, the expected value of $\theta_{13}$ is small
as expected of ${\cal O}(\tilde \alpha)$
in Eqs.~(\ref{ue31}) and (\ref{ue32}).

%%%%%%%%%%%%%%%%%%%%%%%%%%%
\begin{wrapfigure}{r}{7cm}
\begin{center}
\includegraphics[width=7 cm]{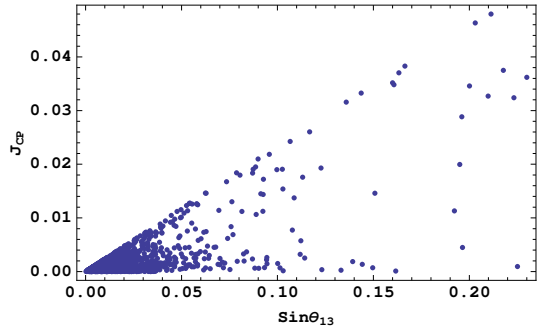}
\caption{The allowed region on  $\sin\theta _{13}$--$J_{CP}$ plane.}
\end{center}
\end{wrapfigure}
%%%%%%%%%%%%%%%%%%%%%%%%%%%%%

It is noted that the main contribution on  $\sin\theta_{13}$ comes from
the next-to-leading term in  Majorana mass matrix $\Delta M_N$  
in Eq.~(\ref{majoranacorr}), which is a few times larger than  
the correction  $\Delta M_D$  of Eq.~(\ref{diraccorr}).

We can estimate the leptonic $CP$ violating measure  $J_{CP}$ since
Yukawa couplings are taken to be complex.
In Figure 4, we show  the  $J_{CP}$ versus $\sin\theta_{13}$. 
 The upper bound of  $J_{CP}$ could be larger than  $0.01$,
which may encourage the measurement  of the $CP$ violation in the
future neutrino oscillation experiments.
However, the probable value of   $J_{CP}$ is much smaller than $0.01$.
We have  estimated the  mean value and its  standard deviation 
of  $J_{CP}$.
The predicted mean value of  $J_{CP}$ is $2.1\times 10^{-3}$ and 
its standard deviation is $4.5\times10^{-3}$.

\subsection{Quark sector}
%%%%%%%%%%%%%%%%%%%%%%%%%%%%%%%%%%%%%%%%%%%%%%%%%%%%%%%%
%%%%%%%%%%        Quark sector      %%%%%%%%%%%%%%%%%%%%
%%%%%%%%%%%%%%%%%%%%%%%%%%%%%%%%%%%%%%%%%%%%%%%%%%%%%%%%

Let us discuss the quark sector.
For down-type quarks,  we can write the superpotential as follows:
\begin{align}
w_d &= y_1\left [\frac{ 1}{\sqrt 2}(s^c\chi _9- b^c \chi _{10}) q_1+\frac{1}{\sqrt 6}(-2d^c \chi _8+s^c\chi _9+b^c\chi _{10})q_2\right ]
h_{45}\Theta ^{\ell }/(\Lambda \bar \Lambda ^{\ell }) \nonumber \\
&\ + y_2( d^c\chi _{11}+ s^c \chi _{12}+  b^c\chi _{13})q_3 h_d/\Lambda .
\end{align}
Since the vacuum alignment is fixed in the lepton sector 
as seen in Eq.~(\ref{alignment1}), 
the down-type quark mass matrix at the leading order is given as
\begin{equation}
M_d = v_d\begin{pmatrix}
            0 & 0 & 0 \\ 
            \bar y_1\lambda ^\ell \alpha _9/\sqrt 2 & \bar y_1\lambda ^\ell \alpha _9/\sqrt 6 & 0 \\
            0 & 0 & y_2\alpha _{13}
         \end{pmatrix},
\end{equation}
where we denote $\bar y_1v_d=y_1v_{45}$.
Then, we have
\begin{equation}
M_d^\dagger M_d = v_d^2\begin{pmatrix}
                          \frac{1}{2}|\bar y_1\lambda ^\ell \alpha _9|^2 & \frac{1}{2\sqrt 3}|\bar y_1\lambda ^\ell \alpha _9|^2 & 0 \\
                          \frac{1}{2\sqrt 3}|\bar y_1\lambda ^\ell \alpha _9|^2 & \frac{1}{6}|\bar y_1\lambda ^\ell \alpha _9|^2 & 0 \\
                          0 & 0 & |y_2|^2\alpha _{13}^2
                       \end{pmatrix}.
\end{equation}
%After rotating by $\theta_{12}^d=60^\circ$, $M_d^\dagger M_d$ turns to be
%\begin{equation}
% v_d^2\begin{pmatrix}
%         0 & 0 & 0 \\
%         0 & \frac{2}{3}|y_1\alpha _{10}+\bar y_1\alpha _{10}^\prime |^2 & 0 \\
%         0 & 0 & |y_2|^2\alpha _{14}^2
%      \end{pmatrix}.
%\end{equation}
This matrix can be diagonalized  by the orthogonal matrix $U_d^{(0)}$ as
\begin{align}
U_d^{(0)} = \begin{pmatrix}
            \cos 60^\circ & \sin 60^\circ & 0 \\
            -\sin 60^\circ & \cos 60^\circ & 0 \\
            0 & 0 & 1
         \end{pmatrix}.
\label{Ud}
\end{align}
The down-type quark masses  are given as 
\begin{align}
&m_d^2=0\ ,
\quad  
m_s^2=\frac{2}{3}|\bar y_1\lambda ^\ell \alpha _9|^2v_d^2\ ,
\quad 
m_b^2=|y_2|^2\alpha _{13}^2v_d^2\ ,
\label{downmass}
\end{align}
which correspond to  ones of  charged lepton masses in Eq.~(\ref{chargemass}). 
The down quark mass vanishes as well as the electron mass, 
however tiny masses appear at  the next-to-leading order. 

%%%%%%%%%%%%%%%%%%%%%%%%%%%%%%%%%%%%%%%%%%%%%%%%%%%%%%%
%%%%%%%%%%%%  next-to-leading Down-type Quarks %%%%%%%%%%%%%%%
%%%%%%%%%%%%%%%%%%%%%%%%%%%%%%%%%%%%%%%%%%%%%%%%%%%%%%%
The down-type quark mass matrix including the next-to-leading order is
\begin{equation}
M_d\simeq
\begin{pmatrix}
\bar\epsilon _{11} & \bar\epsilon _{21} & \bar\epsilon _{31} \\
\frac{\sqrt 3m_s}{ 2}+\bar\epsilon_{12} & \frac{m_s}{2}+\bar\epsilon_{22} 
& \bar\epsilon _{32} \\
\bar\epsilon _{13} & \bar\epsilon _{23} & m_b+\bar\epsilon_{33}
\end{pmatrix},
\label{nextleading}
\end{equation}
where $\bar\epsilon_{ij}$'s are given 
by replacing  $\bar y_{\Delta_i}$  with  
 $-\bar y_{\Delta_i}/3\ (i=c_1,c_2,d,f)$ in Eq.~(\ref{correction}),
and  $m_s$ and $m_b$ are given in  Eq.~(\ref{downmass}).

In order to get the left-handed mixing, we estimate $M_d^\dagger M_d$ as
\begin{footnotesize}
\begin{align}
&M_d^\dagger M_d\simeq  \nonumber\\
&\begin{pmatrix}
|\frac{\sqrt 3m_s}{2}+\bar \epsilon _{12}|^2+|\bar \epsilon _{11}|^2 +|\bar \epsilon _{13}|^2
&(\frac{\sqrt 3}{2}m_s+\bar \epsilon _{12}^*)(\frac{1}{2}m_s+\bar \epsilon _{22})
+\bar \epsilon _{11}^*\bar \epsilon _{21}+\bar \epsilon _{13}^*\bar \epsilon _{23}
&\bar \epsilon _{13}^*m_b \\
(\frac{\sqrt 3}{2}m_s+\bar \epsilon _{12})(\frac{1}{2}m_s+\bar \epsilon _{22}^*)
+\bar \epsilon _{11}\bar \epsilon _{21}^*+\bar \epsilon _{13}\bar \epsilon _{23}^*
&|\frac{m_s}{2}+\bar \epsilon _{22}|^2+|\bar \epsilon _{21}|^2 +|\bar \epsilon _{23}|^2
& \bar \epsilon _{23}^*m_b \\
\bar \epsilon _{13}m_b & \bar \epsilon _{23}m_b & m_b^2
\end{pmatrix},
\end{align}
\end{footnotesize}
By rotating the matrix $M_d^\dagger M_d$ with the mixing matrix $U_d^{(0)}$  
in Eq.~(\ref{Ud}), we have
\begin{equation}
{U_d^{(0)}}^\dagger M_d^\dagger M_dU_d^{(0)}\simeq 
\begin{pmatrix}
m_d^2 & {\mathcal O}(m_dm_s) & \frac{1}{2}(\bar \epsilon _{13}^*-\sqrt 3\bar \epsilon _{23}^*)m_b \\
{\mathcal O}(m_d m_s) & m_s^2 & \frac{1}{2}(\sqrt 3\bar \epsilon _{13}^*+\bar \epsilon _{23}^*)m_b \\
\frac{1}{2}(\bar \epsilon _{13}-\sqrt 3\bar \epsilon _{23})m_b & \frac{1}{2}(\sqrt 3\bar \epsilon _{13}+\bar \epsilon _{23})m_b & m_b^2
\end{pmatrix}.
\label{Md22}
\end{equation}
Then, we get mixing angles $\theta _{12}^d,\ \theta _{13}^d,\ \theta _{23}^d$ 
in the mass matrix of  Eq.~(\ref{Md22}) as
\begin{eqnarray}
\theta _{12}^d=\mathcal{O}\left (\frac{m_d}{m_s}\right )
=\mathcal{O}\left (0.05\right ),\ \  
\theta _{13}^d=\mathcal{O}\left (\frac{m_d}{m_b}\right )
=\mathcal{O}\left (0.005\right ),\ \ 
\theta _{23}^d=\mathcal{O}\left (\frac{m_d}{m_b}\right )
=\mathcal{O}\left (0.005\right ),
\label{dangles}
\end{eqnarray}
where $CP$ violating phases are neglected.

%%%%%%%%%%%%%%%%%%%%%%%%%%%%%%%%%%%%%%%%%
%%%%%%%%%%%%%%%% Up Quarks %%%%%%%%%%%%%%
%%%%%%%%%%%%%%%%%%%%%%%%%%%%%%%%%%%%%%%%%
Let us discuss the up-type quark sector. The superpotential  respecting 
 $S_4 \times Z_4\times U(1)_{FN}$  is given as
\begin{align}
w_u=y_1^u\left [(u^c\chi _1+c^c\chi _2)q_3+t^c(q_1\chi _1+q_2\chi _2)\right ] h_u/\Lambda  +y_2^u t^c q_3h_u \ .
\end{align}
We denote  their VEVs as follows:
\begin{equation}
\langle (\chi _1,\chi _2)\rangle =(u_1,u_2)\ .
\end{equation}
Then, we obtain the mass matrix for up-type quarks is given as
\begin{equation}
M_u = v_u\begin{pmatrix}
0 & 0 & y_1^u\alpha _1 \\ 
0 & 0 & y_1^u\alpha _2 \\
y_1^u\alpha _1 & y_1^u\alpha _2 & y_2^u
\end{pmatrix}.
\label{Mu}
\end{equation}
%%%%%%%%%%%%%%%%%%

 The next-to-leading terms of the superpotential are also important
for the prediction of the $CP$ violation in the quark sector.
The relevant superpotential 
 is given at the next-to-leading order  as 
\begin{align}
\Delta w_u&=y_{\Delta _a}^u(T_1,T_2)\otimes (T_1,T_2)\otimes (\chi _1,\chi _2)\otimes (\chi _1,\chi _2)\otimes H_{5}/\Lambda ^2 \nonumber \\
&\ +y_{\Delta _b}^u(T_1,T_2)\otimes (T_1,T_2)\otimes \chi_{14}\otimes \chi_{14}\otimes H_{5}/\Lambda ^2 \nonumber \\
&\ +y_{\Delta _c}^uT_3 \otimes T_3\otimes (\chi _8,\chi _9,\chi _{10})\otimes (\chi _8,\chi _9,\chi _{10})\otimes H_{5}/\Lambda ^2  .
\end{align}
Then, the next-to-leading mass matrix becomes 
\begin{footnotesize}
%\begin{small}
\begin{eqnarray}
\hskip -1.5cm &&\Delta M_u= v_u \times  \nonumber\\
\hskip -1.5cm&& \left ( 
\begin{matrix}
y_{\Delta _{a_1}}^u(\alpha _1^2+\alpha _2^2)+y_{\Delta _{a_2}}^u(\alpha _1^2-\alpha _2^2)+y_{\Delta _{b}}^u\alpha_{14}^2 & y_{\Delta _{a_2}}^u\alpha _1 \alpha _2 & 0 \cr
y_{\Delta _{a_2}}^u\alpha _1 \alpha _2 & y_{\Delta _{a_1}}^u(\alpha _1^2+\alpha _2^2)-y_{\Delta _{a_2}}^u(\alpha _1^2-\alpha _2^2)+ y_{\Delta _{b}}^u\alpha _{14}^2 & 0 \cr
0 & 0 & y_{\Delta _{c}}^u\alpha _9^2 \cr
\end{matrix}
\right ) , 
\end{eqnarray}
%\end{small}
\end{footnotesize}

\noindent which is added to the leading up-type quark one of Eq.~(\ref{Mu}). 
In order to get the realistic quark mixing, we take the alignment
\begin{equation}
\alpha _1=\alpha _2 ,
\end{equation}
%%%%%%%%%%%%%%%%%%
then, the mass matrix of the up-type quarks is 
\begin{equation}
 M_u=v_u
\begin{pmatrix}
2y_{\Delta _{a_1}}^u\alpha _1^2+y_{\Delta _{b}}^u\alpha_{14}^2 & y_{\Delta _{a_2}}^u\alpha _1^2 & y_1^u\alpha _1 \\
y_{\Delta _{a_2}}^u\alpha _1^2 & 2y_{\Delta _{a_1}}^u\alpha _1^2+y_{\Delta _{b}}^u\alpha_{14}^2 & y_1^u\alpha _1 \\
y_1^u\alpha _1 & y_1^u\alpha _1 & y_2^u+y_{\Delta _c}^u\alpha _9^2
\end{pmatrix}.
\end{equation}
After rotating $M_u$ by the orthogonal matrix $U_u^{(0)}$ as 
\begin{eqnarray}
U_u^{(0)}=
\begin{pmatrix}
\cos45^\circ &\sin45^\circ &0\\ 
-\sin45^\circ &\cos45^\circ &0\\ 
0 &0 &1\\ 
\end{pmatrix},
\end{eqnarray}
we get
\begin{equation}
\hat M_u=U_u^\dagger M_u U_u =
v_u\begin{pmatrix}
(2y_{\Delta _{a_1}}^u-y_{\Delta _{a_2}}^u)\alpha _1^2+y_{\Delta _{b}}^u\alpha_{14}^2 & 0 & 0 \\
0 & (2y_{\Delta _{a_1}}^u+y_{\Delta _{a_2}}^u)\alpha _1^2+y_{\Delta _{b}}^u\alpha_{14}^2 & \sqrt 2y_1^u\alpha _1 \\
0 & \sqrt 2y_1^u\alpha _1 & y_2^u+y_{\Delta _c}^u\alpha _9^2
\end{pmatrix}.
\label{umass}
\end{equation}

We take a phase convention  in which $(1,1)$ and  $(3,3)$ elements
in Eq.~(\ref{umass}) are real.
It is found that  the magnitude of $(2,2)$ element is much smaller than 
that of  $(2,3)$ and  $(3,3)$ elements.
 In the limit of neglecting the   $(2,2)$ element,
the mass matrix $\hat M_u$ is taken to be real 
since  other phases  can be removed  by the phase matrix $P$
 \begin{equation}
  P =\begin{pmatrix}
     1 & 0& 0 \\ 0& e^{-i\rho}& 0\\ 0 & 0 & 1
             \end{pmatrix}.
\end{equation}
The matrix is diagonalized by the orthogonal transformation
 as   $V_u^T \hat M_u V_u$, where
\begin{equation}
V_u= 
\begin{pmatrix}
1 & 0  & 0 \\
0 &  r_t & r_c \\
0 & -r_c &  r_t
\end{pmatrix},
\qquad r_c=\sqrt \frac{m_c}{m_c+m_t}\ ,\qquad r_t=\sqrt \frac{m_t}{m_c+m_t}\ ,
\end{equation}
in which mass eigenvalues of up-type quarks  are given as
\begin{eqnarray}
&&m_u= \left [(2y_{\Delta _{a_1}}^u-y_{\Delta _{a_2}}^u)\alpha _1^2+y_{\Delta _{b}}^u\alpha_{14}^2\right ]v_u ,\nonumber\\
&&m_c\simeq \frac{y_2^u\left [(2y_{\Delta _{a_1}}^u+y_{\Delta _{a_2}}^u)\alpha _1^2+y_{\Delta _{b}}^u\alpha_{14}^2\right ]-2{y_1^u}^2\alpha _1^2}{y_2^u}v_u,
\qquad
m_t\simeq y_2^uv_u.
\label{uptypequarkmass}
\end{eqnarray}

%%%%%%%%%%%%%%%%%%%%%%%%%%%%%%%%%
%%%%%%%%%%%%%%%%%%%%%%%%%%%%%%%%%

Now we can   discuss  the CKM matrix.
Mixing matrices of up- and down-type quarks are summarized as
\begin{eqnarray}
&&U_u \simeq U_u^{(0)}PV_u = 
\begin{pmatrix}
\cos 45^\circ & \sin 45^\circ & 0 \\
-\sin 45^\circ & \cos 45^\circ & 0 \\
0 & 0 & 1
\end{pmatrix}
\begin{pmatrix}
1 & 0& 0 \\
0 & e^{-i\rho} & 0\\
0 & 0 & 1
\end{pmatrix}
\begin{pmatrix}
1 & 0  & 0 \\
0 &  r_t & r_c \\
0 & -r_c &  r_t
\end{pmatrix},
\nonumber\\
&&U_d \simeq \begin{pmatrix}
                         \cos 60^\circ & \sin 60^\circ & 0 \\
                        -\sin 60^\circ & \cos 60^\circ & 0 \\
                            0 & 0 & 1
                                                 \end{pmatrix}
\begin{pmatrix}
1 & \theta_{12}^d & \theta_{13}^d \\ 
-\theta _{12}^d-\theta _{13}^d\theta _{23}^d  & 1 & \theta _{23}^d \\
-\theta _{13}^d+\theta _{12}^d\theta _{23}^d  & -\theta _{23}^d-\theta _{12}^d\theta _{13}^d & 1 \\
\end{pmatrix}.
\end{eqnarray}
Therefore, the CKM matrix at the GUT scale can be written as 
\begin{align}
V^{0}= U_u^\dagger U_d
\simeq & 
\begin{pmatrix}     
1 & 0 & 0 \\
0 & r_t & -r_c \\
0 & r_c & r_t
\end{pmatrix}
\begin{pmatrix}
1 & 0 & 0 \\
0 & e^{i\rho} & 0 \\
0 & 0 & 1
\end{pmatrix}
\nonumber \\
&\times 
\begin{pmatrix}
\cos 15^\circ & \sin 15^\circ & 0 \\
-\sin 15^\circ & \cos 15^\circ & 0 \\
0 & 0 & 1
\end{pmatrix} 
\begin{pmatrix}
1 & \theta_{12}^d & \theta_{13}^d \\ 
-\theta _{12}^d-\theta _{13}^d\theta _{23}^d  & 1 & \theta _{23}^d \\
-\theta _{13}^d+\theta _{12}^d\theta _{23}^d  & -\theta _{23}^d-\theta _{12}^d\theta _{13}^d & 1 \\
\end{pmatrix}.
\end{align}
In the following, we suppose  that  $\theta_{12}^d$, 
$\theta_{13}^d$, and $\theta_{23}^d$ are real.
Then, nine CKM matrix elements at the GUT scale are  expressed  as
\begin{equation}
\begin{split}
V_{ud}^{0}
&\simeq \cos 15^\circ -(\theta _{12}^d+\theta _{13}^d\theta _{23}^d)\sin 15^\circ ,
\\
V_{us}^{0}
&\simeq \theta _{12}^d\cos 15^\circ +\sin 15^\circ ,
\\
V_{ub}^{0}
&\simeq \theta _{13}^d\cos 15^\circ +\theta _{23}^d\sin 15^\circ ,
\\
V_{cd}^{0}
&\simeq -r_te^{i\rho }\sin 15^\circ -r_t(\theta _{12}^d+\theta _{13}^d\theta _{23}^d)e^{i\rho }\cos 15^\circ +r_c(\theta _{13}^d-\theta _{12}^d\theta _{23}^d)\ ,
\\
V_{cs}^{0}
&\simeq -r_t\theta _{12}^de^{i\rho }\sin 15^\circ +r_te^{i\rho }\cos 15^\circ +r_c(\theta _{23}^d+\theta _{12}^d\theta _{13}^d)\ ,
\\
V_{cb}^{0}
&\simeq -r_t\theta _{13}^de^{i\rho }\sin 15^\circ +r_t\theta _{23}^de^{i\rho }\cos 15^\circ -r_c\ ,
\\
V_{td}^{0}
&\simeq -r_c\sin 15^\circ e^{i\rho }-r_c(\theta _{12}^d+\theta _{13}^d\theta _{23}^d)e^{i\rho }\cos 15^\circ +r_t(-\theta _{13}^d+\theta _{12}^d\theta _{23}^d) \ ,
\\
V_{ts}^{0}
&\simeq -r_c\theta _{12}^d\sin 15^\circ e^{i\rho }+r_ce^{i\rho }\cos 15^\circ -r_t(\theta _{23}^d+\theta _{12}^d\theta _{23}^d) \ ,
\\
V_{ts}^{0}
&\simeq -r_c\theta _{12}^d\sin 15^\circ e^{i\rho }+r_c\theta _{23}^de^{i\rho }\cos 15^\circ +r_t \ .
\end{split}
\end{equation}

Putting typical masses  at the GUT scale 
 $m_u=1.04\times 10^{-3}$ GeV, $m_c=302\times 10^{-3}$ GeV, and $m_t=129$ GeV~\cite{Fusaoka:1998vc}, 
we can write  CKM matrix elements in terms of $\rho$ and $\theta _{ij}^d$
at the GUT scale.
We should take account the  renormalization effect
in order to get the  CKM matrix elements   at the electroweak (EW) scale.
We use a  simple formula of the renormalization 
for the CKM matrix in Ref.~\cite{JuarezWysozka:2002kx}.
%\begin{equation}
%|V_{\text{CKM}}^{\text{exp}}|\simeq 
%\begin{pmatrix}
%0.974 & 0.226 & 0.0036 \\ %3.59\times 10^{-3} \\ 
%0.226 & 0.973 & 0.042  \\ %4.15\times 10^{-2} \\
%0.0087& 0.041 & 0.999%8.74\times 10^{-3} & 4.07\times 10^{-2} & 0.999
%\end{pmatrix}.
%\end{equation}
%With renormalization group equation, 
Then, the CKM matrix at the GUT scale~\cite{JuarezWysozka:2002kx} becomes 

\begin{eqnarray}
\begin{pmatrix}
V_{ud}^0   & V_{us}^0  & V_{ub}^0/h(t)  \\ 
V_{cd}^0   & V_{cs}^0  & V_{cb}^0/h(t)  \\ 
V_{td}^0/h(t)   & V_{ts}^0/h(t)  & V_{tb}^0  \\ 
 \end{pmatrix}_\text{EW},
\end{eqnarray}
at the EW scale.
In the case of the GUT scale $10^{16}$ GeV, we have 
$h(t)\simeq 1.05$. 
Putting  central values of the observed  CKM matrix elements in
the particle data group~\cite{PDG}, 
i.e. $|V_{us}|=0.2257$, $|V_{ub}|=0.00359$, $|V_{cb}|=0.0415$, 
and $|V_{td}|=0.00874$, we obtain a parameter set  
\begin{eqnarray}
\rho=123^\circ, \quad
\theta _{12}^d= -0.0340,
\quad
\theta _{13}^d= 0.00626,
\quad
\theta _{23}^d= -0.00880,
\end{eqnarray}
which reproduce the experimental data.
These magnitudes of $\theta_{ij}^d$'s are consistent with
 the ones in Eq.~(\ref{dangles}).

In terms of a phase $\rho$,
we can  also estimate the magnitude of the $CP$ violation.
Let us calculate the $CP$ violation measure, 
 Jarlskog invariant $J_{CP}$~\cite{Jarlskog:1985ht}, which is given as
\begin{eqnarray}
|J_{CP}|=|\text{Im}\left\{ V_{us} V_{cs}^* V_{ub} V_{cb}^* \right\} 
| \simeq 3.06\times 10^{-5}\ ,
\end{eqnarray}
 where $\rho=123^\circ$ is taken.
Our prediction is  consistent with the experimental value
$J_{CP}=(3.05^{+0.19}_{-0.20})\times 10^{-5}$~\cite{PDG}.

Next, we calculate  three angles of the unitarity triangle,
$\alpha({\rm or}\ \phi_2)$, $\beta({\rm or}\ \phi_1)$, and 
$\gamma({\rm or}\ \phi_3)$,
\begin{eqnarray}
\alpha=
{\rm arg}\left( -\frac{V_{td}V_{tb}^*}{V_{ud}V_{ub}^*}\right ) ,\qquad
\beta=
{\rm arg}\left( -\frac{V_{cd}V_{cb}^*}{V_{td}V_{tb}^*}\right ) ,\qquad
\gamma=
{\rm arg}\left( -\frac{V_{ud}V_{ub}^*}{V_{cd}V_{ccb}^*}\right ) .
\end{eqnarray}
Putting $\rho=123^\circ$, we obtain
$\alpha=89.4^\circ $, $\sin2\beta=0.693$ ($\beta=21.9^\circ $),
and $\gamma=68.7^\circ $, 
which are consistent with experimental values
$\alpha=(88^{+6}_{-5})^\circ$, $\sin 2\beta=0.681\pm 0.025$, 
and  $\gamma=(77^{+30}_{-32})^\circ$~\cite{PDG}.

%%%%%%%%%%%%%%%%%%%%%%%%%%%%%%%%%%%%%%%%%%%%%
%%%%%%%%%%%%%%%%%%%%%%%%%%%%%%%%%%%%%%%%%%%%%
%%%%%%%%%%%%%%%%%%%%%%%%%%%%%%%%%%%%%%%%%%%%%

\section{Magnitudes of VEVs and Alignment}

As seen in previous section, we need relevant  vacuum alignment
 to get the tri-bimaximal mixing of leptons and Cabibbo angle
 of quarks.   The alignment of VEVs is   summarized as
\begin{align}
& (\chi _1, \chi _2)= (1,1), \quad 
 (\chi _3, \chi _4)= (0,1), \nonumber \\
&(\chi _5, \chi _6, \chi _7)=(1,1,1), 
\quad (\chi _8, \chi _9, \chi _{10})=(0,1,0), \quad
(\chi _{11}, \chi_{12}, \chi _{13})=(0,0,1), 
\label{alignment}
\end{align}
where  these magnitudes are given in  arbitrary units.
 
Magnitudes of  $\alpha_i\equiv  \langle \chi_i \rangle / \Lambda$ 
are determined when  the quark and lepton masses are put, except
for $\alpha_{14}$, which appears at the next-to-leading order.
Here, we have fixed
$\ell=1$, $m=1$, and $n=2$ as Frogatt-Nielsen charges, which are discussed in
  Appendix B. 
Then, these are given as
\begin{align}
&\alpha_ 3=\alpha _8=\alpha _{10}=\alpha _{11}=\alpha _{12}=0, \nonumber \\
&\alpha _1=\alpha _2\simeq\sqrt{\frac{m_c}{2\left |y_{\Delta _{a_2}}^u-\frac{{y_1^u}^2}{y_2^u}\right |v_u}}~, 
\nonumber\\
& \alpha _4 = \frac{(y_1^D\lambda )^2(m_3-m_1)m_2M}{6y_2^N{y_2^D}^2m_1m_3\Lambda },
\qquad
\alpha _5 = \alpha _6 = \alpha _7 = \frac{\sqrt{m_2M}}{\sqrt 3y_2^Dv_u}, 
\nonumber \\
& \alpha _9= \frac{m_\mu }{\sqrt 6|\bar{y_1}|\lambda v_d}, 
\qquad \alpha _{13} = \frac{m_\tau }{y_2v_d}\ .
\label{alphas}
\end{align}
where  masses of quarks and leptons are given at the GUT scale.

Putting typical values of quark masses at the GUT scale~\cite{Fusaoka:1998vc},
$M=10^{12}~\ \text{GeV}$, $\lambda=0.1$, 
and $\tan\beta=3$ ($v_d\simeq 55~\text{GeV}$, $v_u\simeq 165~\text{GeV}$)
with of order $1$ for  absolute values of Yukawa couplings,
 we have 
\begin{align}
&\alpha _1 \sim 3\times 10^{-2},
\qquad 
\alpha _4 \sim 10^{-2},
\qquad  
\alpha _5 \sim 10^{-2},
\nonumber \\
\qquad 
&\alpha _9 \sim 5\times 10^{-3},
\qquad 
\alpha _{13}\sim 2\times 10^{-2}.
\label{valuealpha}
\end{align}
Therefore, the magnitudes of all VEVs are almost of order $10^{-2}$.
Hereafter, we denote the averaged value as $\tilde\alpha$.

%%%%%%%%%%%%%%%%%%%%%%%%%%%%%%%%%%%%
%%%%%%%%%%%%%%%%%%%%%%%%%%%%%%%%%%%%

 We can generate the vacuum alignment through $F$-terms by
coupling flavons to driving fields, which carry the $R$ charge $+2$
under $U(1)_R$ symmetry. 

\begin{table}[h]
\begin{tabular}{|c|cccccc|}
\hline
& $(\chi _1,\chi _2)$ & $(\chi _3,\chi _4)$ & $(\chi _5,\chi _6,\chi _7)$ 
& $(\chi _8,\chi _9,\chi _{10})$ & $(\chi _{11},\chi _{12},\chi _{13})$ & $\chi _{14}$ \\ \hline
$SU(5)$ & $1$ & $1$ & $1$ & $1$ & $1$ & $1$ \\
$S_4$ & $\bf 2$ & $\bf 2$ & ${\bf 3}'$ & $\bf 3$ & $\bf 3$ & $\bf 1$ \\
$Z_4$ & $-i$ & $1$ & $-i$ & $-1$ & $i$ & $i$ \\
$U(1)_{FN}$ & $-\ell $ & $-n$ & $0$ & $0$ & $0$ & $-\ell $ \\
$U(1)_R$ & $0$ & $0$ & $0$ & $0$ & $0$ & $0$ \\
\hline
\end{tabular}
\end{table}
\vspace{-0.5cm}
\begin{table}[h]
\begin{tabular}{|c|c||cccc|}
\hline
& $(\chi _{15},\chi _{16},\chi _{17})$ & $\chi _1^0$ & $\chi _2^0$ & $\chi _3^0$ & $(\chi _4^0,\chi _5^0)$ \\ \hline
$SU(5)$ & $1$ & $1$ & $1$ & $1$ & $1$ \\
$S_4$ & $\bf 3$ & $\bf 1$ & $\bf 1$ & $\bf 1$ & $\bf 2$ \\
$Z_4$ & $-1$ & $-1$ & $i$ & $-1$ & $-i$ \\
$U(1)_{FN}$ & $-z$ & $2\ell +n$ & $0$ & $2\ell $ & $z$ \\
$U(1)_R$ & $0$ & $2$ & $2$ & $2$ & $2$ \\
\hline
\end{tabular}
\caption{Assignments of $SU(5)$, $S_4$, $Z_4$, $U(1)_{FN}$, and $U(1)_R$ representations for flavons and driving fields.}
\label{tables4}
\end{table}
  Three  $S_4$ singlets $\chi_1^0$,  $\chi_2^0$, and $\chi_3^0$ and
 one  $S_4$ doublet $(\chi _4^0,\chi _5^0)$ 
for driving fields are required
to obtain relevant  vacuum alignment in our model.
Moreover, an $S_4$ triplet $(\chi _{15},\chi _{16},\chi _{17})$ 
is introduced as additional  flavons\footnote{As far as $z\gg 1$, $(\chi _{15},\chi _{16},\chi _{17})$ do not disturb the result in section 2.}. 
Assignments of flavons and driving fields are summarized in
Table \ref{tables4}.

The $SU(5)\times S_4\times Z_4\times U(1)_{FN}\times U(1)_R$ 
invariant superpotential is given as 
\begin{align}
w' &= \kappa _1\left (\chi _1,\chi _2\right )\otimes \left (\chi _1,\chi _2\right )\otimes \left (\chi _3,\chi _4\right )\otimes \chi _1^0/\Lambda \nonumber \\
&\ +\eta _1\left (\chi _8,\chi _9,\chi _{10}\right )\otimes \left (\chi _{11},\chi _{12},\chi _{13}\right )\otimes \chi _2^0 \nonumber \\
&\ +\eta _2\left (\chi _1,\chi _2\right )\otimes \left (\chi _1,\chi _2\right )\otimes \chi _3^0+\eta _3\chi _{14}\otimes \chi _{14}\otimes \chi _3^0 \nonumber \\
&\ +\eta _4\left (\chi _5,\chi _6,\chi _7\right )\otimes \left (\chi _{15},\chi _{16},\chi _{17}\right )\otimes \left (\chi _4^0,\chi _5^0\right ),
\label{scalar-leading}
\end{align}
which is rewritten as 
\begin{align}
w' &= \kappa _1\left [2\chi _1\chi _2\chi _3+\left (\chi _1^2-\chi _2^2\right )\chi _4\right ]\chi _1^0/\Lambda 
\ + \eta _1\left (\chi _8\chi _{11}+\chi _9\chi _{12}+\chi _{10}\chi _{13}\right )\chi _2^0 \nonumber \\
&\ + \left [\eta _2(\chi _1^2+\chi _2^2)+\eta _3\chi _{14}^2\right ]\chi _3^0 
\ + \frac{1}{\sqrt 2}\eta _4\left (\chi _6\chi _{16}-\chi _7\chi _{17}\right )\chi _4^0 \nonumber \\
&\ + \frac{1}{\sqrt 6}\eta _4\left (-2\chi _5\chi _{15}+\chi _6\chi _{16}+\chi _7\chi _{17}\right )\chi _5^0 \ .
\end{align}
Then the scalar potential is given  as 
\begin{align}
V &= \left |\frac{\kappa _1}{\Lambda}
\left [2\chi _1\chi _2\chi _3+\left (\chi _1^2-\chi _2^2\right )\chi _4\right ] \right |^2 
 + \left |\eta _1\left (\chi _8\chi _{11}+\chi _9\chi _{12}+\chi _{10}\chi _{13}\right )\right |^2 \nonumber \\
&\ + \left |\eta _2(\chi _1^2+\chi _2^2)+\eta _3\chi _{14}^2\right |^2 \nonumber 
+ \left |\frac{1}{\sqrt 2}\eta _4\left (\chi _6\chi _{16}-\chi _7\chi _{17}\right )\right |^2 \nonumber \\
&\ + \left |\frac{1}{\sqrt 6}\eta _4\left (-2\chi _5\chi _{15}+\chi _6\chi _{16}+\chi _7\chi _{17}\right )\right |^2 \ .
\end{align}
Therefore conditions to realize  the potential minimum ($V=0$) 
are given  as 
\begin{align}
\kappa _1\left [2\chi _1\chi _2\chi _3+\left (\chi _1^2-\chi _2^2\right )\chi _4\right ]/\Lambda &= 0, \nonumber \\
\eta _1\left (\chi _8\chi _{11}+\chi _9\chi _{12}+\chi _{10}\chi _{13}\right ) &= 0, \nonumber \\
\eta _2(\chi _1^2+\chi _2^2)+\eta _3\chi _{14}^2 &= 0, \nonumber \\
\frac{1}{\sqrt 2}\eta _4\left (\chi _6\chi _{16}-\chi _7\chi _{17}\right ) &= 0 ,\nonumber \\
\frac{1}{\sqrt 6}\eta _4\left (-2\chi _5\chi _{15}+\chi _6\chi _{16}+\chi _7\chi _{17}\right ) &= 0 \ ,
\end{align}
where $\chi _i$'s are regarded as VEVs. 
One of the  solution  which satisfies  these conditions is obtained as
\begin{align}
\chi _1=\chi _2,\quad &\chi _3=0,\quad \chi _5=\chi _6=\chi _7,\quad \chi _8=\chi _{10}=\chi _{11}=\chi _{12}=0,\nonumber \\
&\chi _{14}^2=-\frac{2\eta _2}{\eta _3}\chi _1^2,\quad \chi _{15}=\chi _{16}=\chi _{17}\ .
\label{vacuum-alignment}
\end{align}
Therefore we obtain the desired  alignment  of VEVs in Eq.~(\ref{alignment}).
%as follows: 
%\begin{align}
%&(\chi _1, \chi _2)= (1,1), \quad 
%(\chi _3, \chi _4)= (0,1), \quad 
%(\chi _5, \chi _6, \chi _7)=(1,1,1), \nonumber \\
%&(\chi _8, \chi _9, \chi _{10})=(0,1,0), \quad
%(\chi _{11}, \chi_{12}, \chi _{13})=(0,0,1), \quad 
%(\chi _{15}, \chi_{16}, \chi _{17})=(1,1,1)\ .
%\label{vacuum-alignment}
%\end{align}
 Next-to-leading couplings of  flavons and  driving fields
could  shift these alignments. Detail discussions are  given
in Appendix C.

%%%%%%%%%%%%%%%%%%%%%%%%%%%%%%%%%%%%%%%%%%%%%%%%%%%%%%%%%%%%%%%
%%%%%%%%%%%%%%%%%%%%%%%%%%%%%%%%%%%%%%%%%%%%%%%%%%%%%%%%%%%%%%%
%%%%%%%%%%%%%%%%%%%%%%%%%%%%%%%%%%%%%%%%%%%%%%%%%%%%%%%%%%%%%%%

\section{Soft SUSY breaking terms}

We have already discussed  SUSY breaking terms
 i.e., 
sfermion masses and scalar trilinear couplings
 in the $D_4$ flavor model~\cite{Ishimori:2008gp,Ishimori:2008ns}
, the $A_4$ flavor model~\cite{Hayakawa:2009va},
and the  $\Delta(54)$ flavor model~\cite{Ishimori:2008uc,Ishimori:2009ew}.
In this section, 
we study SUSY breaking terms
in the framework of $S_4 \times Z_4 \times U(1)_{FN}$.
We consider the gravity mediation within the 
framework of supergravity theory.
We assume that 
non-vanishing $F$-terms of gauge and flavor singlet (moduli) fields $Z$ 
and gauge singlet fields $\chi_i$ $(i=1,\cdots,14)$ 
contribute to the SUSY breaking.
Their $F$-components are written as 
\begin{eqnarray}
F^{\Phi_k}= - e^{ \frac{K}{2M_p^2} } K^{\Phi_k \bar{I} } \left(
  \partial_{\bar{I}} \bar{W} + \frac{K_{\bar{I}}} {M_p^2} \bar{W} \right) ,
\label{eq:F-component}
\end{eqnarray}
where $K$ denotes the K\"ahler potential, $K_{\bar{I}J}$ denotes 
second derivatives by fields, 
i.e. $K_{\bar{I}J}={\partial}_{\bar{I}} \partial_J K$
and $K^{\bar{I}J}$ is its inverse. 
Here the fields ${\Phi_k}$ correspond to the moduli fields $Z$ and 
gauge singlet fields $\chi_i$.
The VEVs of $F_{\Phi_k}/\Phi_k$  are estimated as 
$\langle F_{\Phi_k}/ \Phi_k \rangle = {\cal O}(m_{3/2})$, where
$m_{3/2}$ denotes the gravitino mass, which is obtained as 
$m_{3/2}= \langle e^{K/2M_p^2}W/M_p^2 \rangle$.

\subsection{Soft scalar masses of slepton sector}

First, let us study soft scalar masses.
Within the framework of supergravity theory,
soft scalar mass squared is obtained as~\cite{Kaplunovsky:1993rd}
\begin{eqnarray}
m^2_{\bar{I}J} K_{{\bar{I}J}}= m_{3/2}^2K_{{\bar{I}J}} 
+ |F^{\Phi_k}|^2 \partial_{\Phi_k}  
\partial_{  \bar{\Phi_k} }  K_{\bar{I}J}-
|F^{\Phi_k}|^2 \partial_{\bar{\Phi_k}}  K_{\bar{I}L} \partial_{\Phi_k}  
K_{\bar{M} J} K^{L \bar{M}}.
\label{eq:scalar}
\end{eqnarray}
The invariance under the  $S_4 \times Z_4 \times U(1)_{FN}$
flavor symmetry 
as well as the gauge invariance requires the following form 
of the K\"ahler potential as %of $L_i$ and $R_i$ $(i=e,\mu,\tau)$
\begin{equation}
K = Z^{(L)}(\Phi )\sum_{i=e,\mu,\tau } |L_i|^2 + 
Z_{(1)}^{(R)}(\Phi )\sum_{i=e,\mu }|R_i|^2 +Z_{(2)}^{(R)}(\Phi )|R_\tau |^2, 
\label{eq:Kahler}
\end{equation}
at the lowest level, where $Z^{(L)}(\Phi)$ and $Z_{(1),(2)}^{(R)}(\Phi)$ are 
arbitrary functions of the singlet fields $\Phi$.
By use of Eq.~(\ref{eq:scalar}) with 
the K\"ahler potential in Eq.~(\ref{eq:Kahler}), 
we obtain the following matrix form 
of soft scalar masses squared for left-handed and 
right-handed charged sleptons,
\begin{eqnarray}
(m_{\tilde L}^2)_{ij} =
\left(
  \begin{array}{ccc}
m_{L}^2   &  0 &  0 \\ 
0   & m_{L}^2  & 0  \\ 
0   &  0  & m_{L}^2   \\ 
\end{array} \right ),
\qquad
(m_{\tilde R}^2)_{ij} 
 = 
\left(
  \begin{array}{ccc}
m_{R(1)}^2   &  0 &  0 \\ 
0   &  m_{R(1)}^2 & 0  \\ 
0   & 0   & m_{R(2)}^2   \\ 
\end{array} \right ).
\label{eq:soft-mass-1}
\end{eqnarray}
That is, three left-handed slepton masses are degenerate, and 
two right-handed slepton masses  are degenerate.
These predictions  would be obvious because 
the left-handed sleptons  form  a triplet of  $S_4$, 
and the right-handed slepton form  a doublet and a singlet of $S_4$.
These  predictions hold exactly before $S_4\times Z_4\times U(1)_{FN}$
is broken, 
but its breaking gives next-to-leading terms in the slepton mass matrices.

Next, we study effects due to  $S_4 \times Z_4\times U(1)_{FN}$  breaking 
by $\chi_i$.
That is, we estimate corrections to the K\"ahler potential 
including  $\chi_i$.
Since each VEV is taken as the same order, 
the breaking scale can be characterized by the 
average of VEVs, such as  $\tilde\alpha\Lambda$.

In our model, the right-handed charged leptons $(R_e^c,R_\mu^c)$ are assigned to ${\bf 2}$ and 
its conjugate representation is itself ${\bf 2}$. 
Similarly, the left-handed charged leptons 
$(L_e,L_\mu,L_\tau)$ are assigned to ${\bf 3}$ and 
its conjugation is ${\bf 3}$.  Therefore,
for left-handed sector,  higher dimensional terms are given as
\begin{align}
\Delta K_L 
&= \sum _{i=1,3} Z_{\Delta _{a_i}}^{(L)}(\Phi )
 (L_e,L_\mu ,L_\tau ) \otimes (L_e^c,L_\mu ^c,L_\tau ^c)
\otimes (\chi _i,\chi _{i+1})\otimes (\chi _i^c,\chi _{i+1}^c)/\Lambda ^2
\nonumber \\
&\ +\sum _{i=5,8,11} Z_{\Delta _{b_i}}^{(L)}(\Phi )
 (L_e,L_\mu ,L_\tau ) \otimes (L_e^c,L_\mu ^c,L_\tau ^c)
\otimes (\chi _i,\chi _{i+1},\chi _{i+2})\otimes (\chi _i^c,\chi _{i+1}^c,\chi _{i+2}^c)/\Lambda ^2
\nonumber \\
&\ + Z_{\Delta _c}^{(L)}(\Phi )
 (L_e,L_\mu ,L_\tau ) \otimes (L_e^c,L_\mu ^c,L_\tau ^c)
\otimes \chi _{14}\otimes \chi _{14}^c/\Lambda ^2
\nonumber \\
&\ +Z_{\Delta _d}^{(L)}(\Phi )
 (L_e,L_\mu ,L_\tau ) \otimes (L_e^c,L_\mu ^c,L_\tau ^c)
\otimes \Theta \otimes \Theta ^c/\bar \Lambda ^2.
\end{align}
For example, 
higher dimensional terms 
including $(\chi _1,\chi_2)$ and 
 $(\chi _5, \chi _6 ,\chi_7)$ are explicitly written  as
\begin{align}
\Delta K_L^{\left [\chi _1,\chi _5\right ]} &
= Z_{\Delta _{a_1}}^{(L)}(\Phi )\left [\frac{\sqrt 2|\chi _1|^2}{\Lambda ^2}(|L_\mu |^2-|L_\tau |^2)\right ] \nonumber \\
&\ +Z_{\Delta _{b_5}}^{(L)}(\Phi )\left [\frac{2|\chi _5|^2}{\Lambda ^2}
(L_\mu L_\tau ^\ast +L_\tau L_\mu ^\ast +L_eL_\tau ^\ast +L_\tau L_e^\ast +L_eL_\mu ^\ast +L_\mu L_e^\ast )\right ].
\end{align}

When  we take into account  the corrections from all $\chi_i \chi_j^*$ 
to the K\"ahler potential, 
the soft scalar masses squared for left-handed charged sleptons 
have the following corrections, 
\begin{equation}
(m_{\tilde L}^2)_{ij}=\begin{pmatrix}
                              m_L^2+\mathcal{O}(\tilde \alpha ^2m_{3/2}^2) & \mathcal{O}(\tilde \alpha ^2m_{3/2}^2) & \mathcal{O}(\tilde \alpha ^2m_{3/2}^2) \\
                              \mathcal{O}(\tilde \alpha ^2m_{3/2}^2) & m_L^2+\mathcal{O}(\tilde \alpha ^2m_{3/2}^2) & \mathcal{O}(\tilde \alpha ^2m_{3/2}^2) \\
                              \mathcal{O}(\tilde \alpha ^2m_{3/2}^2) & \mathcal{O}(\tilde \alpha ^2m_{3/2}^2) & m_L^2+\mathcal{O}(\tilde \alpha ^2m_{3/2}^2)
                           \end{pmatrix},
\end{equation}
where $\tilde\alpha$ is a linear combination of $\alpha_i$'s.

For right-handed sector, 
higher dimensional terms are given as
\begin{align}
\Delta K_R 
&= \sum _{i=1,3} Z_{\Delta _{a_i}}^{(R)}(\Phi )
 (R_e,R_\mu ) \otimes (R_e^c,R_\mu ^c)
\otimes (\chi _i,\chi _{i+1})\otimes (\chi _i^c,\chi _{i+1}^c)/\Lambda ^2
\nonumber \\
&\ +\sum _{i=5,8,11} Z_{\Delta _{b_i}}^{(R)}(\Phi )
 (R_e,R_\mu ) \otimes (R_e^c,R_\mu ^c) 
\otimes (\chi _i,\chi _{i+1},\chi _{i+2})\otimes (\chi _i^c,\chi _{i+1}^c,\chi _{i+2}^c)/\Lambda ^2
\nonumber \\
&\ + Z_{\Delta _c}^{(R)}(\Phi )
 (R_e,R_\mu ) \otimes (R_e^c,R_\mu ^c) 
\otimes \chi _{14}\otimes \chi _{14}^c/\Lambda ^2
\nonumber \\
&\ + Z_{\Delta _d}^{(R)}(\Phi )
 (R_e,R_\mu ) \otimes R_\tau ^c 
\otimes (\chi _1,\chi _2)/\Lambda ^2
+ Z_{\Delta _e}^{(R)}(\Phi )
 (R_e^c,R_\mu ^c ) \otimes R_\tau 
\otimes (\chi _1^c,\chi _2^c)/\Lambda ^2
\nonumber \\
&\ +\sum_{i=1,3} Z_{\Delta _{f_i}}^{(R)}(\Phi )
 R_\tau \otimes R_\tau ^c
\otimes (\chi _i,\chi _{i+1})\otimes (\chi _i^c,\chi _{i+1}^c)/\Lambda ^2
\nonumber \\
&\ +\sum _{i=5,8,11} Z_{\Delta _{g_i}}^{(R)}(\Phi )
 R_\tau  \otimes R_\tau ^c
\otimes (\chi _i,\chi _{i+1},\chi _{i+2})\otimes (\chi _i^c,\chi _{i+1}^c,\chi _{i+2}^c)/\Lambda ^2
\nonumber \\
&\ + Z_{\Delta _h}^{(R)}(\Phi )
 R_\tau  \otimes R_\tau ^c
\otimes \chi_{14}\otimes \chi_{14}^c/\Lambda ^2
\nonumber \\
&\ + Z_{\Delta _i}^{(R)}(\Phi )
 (R_e,R_\mu ) \otimes (R_e^c,R_\mu ^c) 
\otimes \Theta \otimes \Theta ^c/\bar \Lambda ^2
\nonumber \\
&\ + Z_{\Delta _j}^{(R)}(\Phi )
 R_\tau  \otimes R_\tau ^c
\otimes \Theta \otimes \Theta ^c/\bar \Lambda ^2.
\end{align}

In the same way, right-handed charged sleptons can be written as
\begin{equation}
(m_{\tilde R}^2)_{ij}=\begin{pmatrix}
                              m_{R(1)}^2+\mathcal{O}(\tilde \alpha ^2m_{3/2}^2) & \mathcal{O}(\tilde \alpha ^2m_{3/2}^2) & \mathcal{O}(\alpha _1m_{3/2}^2) \\
                              \mathcal{O}(\tilde \alpha ^2m_{3/2}^2) & m_{R(1)}^2+\mathcal{O}(\tilde \alpha ^2m_{3/2}^2) & \mathcal{O}(\alpha _1m_{3/2}^2) \\
                              \mathcal{O}(\alpha _1m_{3/2}^2) & \mathcal{O}(\alpha _1m_{3/2}^2) & m_{R(2)}^2+\mathcal{O}(\tilde \alpha ^2m_{3/2}^2)
                           \end{pmatrix}.
\label{RR}
\end{equation}
%where we have $\tilde \alpha \sim \mathcal{O}(10^{-2})$, and $\alpha _1=3.0\times 10^{-2}$
%as seen in Eq.~(\ref{valuealpha}).

 In order to estimate the magnitude of FCNC, we move to super-CKM basis
by diagonalizing the charged lepton mass matrix including  next-to-leading terms.
 For the left-handed slepton mass matrix, we obtain
\begin{eqnarray}
(m_{\tilde L}^2)_{ij}^{(SCKM)}&=&U^\dagger_E (m_{\tilde L}^2)_{ij} U_E
\nonumber\\
&\simeq&
\begin{pmatrix}
       m_L^2+\mathcal{O}(\frac{\tilde \alpha ^2}{\lambda^2}m_{3/2}^2)
 &\mathcal{O}(\tilde\alpha^2m_{3/2}^2)&\mathcal{O}(\tilde\alpha^2m_{3/2}^2)\\
            \mathcal{O}(\tilde \alpha ^2m_{3/2}^2) &
 m_L^2+\mathcal{O}(\frac{\tilde \alpha ^2}{\lambda^2}m_{3/2}^2) 
& \mathcal{O}(\tilde \alpha ^2m_{3/2}^2) \\
                              \mathcal{O}(\tilde \alpha ^2m_{3/2}^2) & \mathcal{O}(\tilde \alpha ^2m_{3/2}^2) & m_L^2+\mathcal{O}(\tilde \alpha ^2m_{3/2}^2)
                           \end{pmatrix},
\end{eqnarray}
where we take the left-handed mixing
\begin{equation}
U_E=\begin{pmatrix}
       1 &  \frac{\tilde\alpha}{\lambda} &  \tilde\alpha\\
    -\frac{\tilde\alpha}{\lambda}-\tilde\alpha^2  & 1  &  \tilde\alpha\\
    -\tilde\alpha+ \frac{\tilde\alpha^2}{\lambda} 
& -\tilde\alpha-  \frac{\tilde\alpha^2}{\lambda}& 1
\end{pmatrix}.
\label{UE}
\end{equation}
In Eq.~(\ref{UE}), the unitarity is satisfied up to ${\cal O}(\tilde\alpha^2)$.

 For the right-handed slepton mass matrix, we obtain
\begin{eqnarray}
(m_{\tilde R}^2)_{ij}^{(SCKM)}&=&V^\dagger_E (m_{\tilde R}^2)_{ij} V_E
\nonumber\\
&\simeq&
\begin{pmatrix}
                              m_{R(1)}^2+\mathcal{O}(\tilde \alpha ^2m_{3/2}^2) & \mathcal{O}(\tilde \alpha ^2m_{3/2}^2) & \mathcal{O}(\alpha _1m_{3/2}^2) \\
                              \mathcal{O}(\tilde \alpha ^2m_{3/2}^2) & m_{R(1)}^2+\mathcal{O}(\tilde \alpha ^2m_{3/2}^2) & \mathcal{O}(\alpha _1m_{3/2}^2) \\
                              \mathcal{O}(\alpha _1m_{3/2}^2) & \mathcal{O}(\alpha _1m_{3/2}^2) & m_{R(2)}^2+\mathcal{O}(\tilde \alpha ^2m_{3/2}^2)
                           \end{pmatrix},
\end{eqnarray}
where we take the right-handed mixing as
\begin{equation}
V_E=
\begin{pmatrix}
       \cos 15^\circ &  \sin 15^\circ &  0\\
    -\sin 15^\circ & \cos 15^\circ  &  0\\
    0& 0& 1
\end{pmatrix}
\times
\begin{pmatrix}
       1 &  \frac{\tilde\alpha^2}{\lambda^2} &  \tilde\alpha\\
    -\frac{\tilde\alpha^2}{\lambda^2}-\tilde\alpha^2  & 1  &  \tilde\alpha\\
    -\tilde\alpha+ \frac{\tilde\alpha^3}{\lambda^2} 
& -\tilde\alpha-  \frac{\tilde\alpha^3}{\lambda^2}& 1
\end{pmatrix}.
\label{VE}
\end{equation}
In Eq.~(\ref{VE}), the unitarity is satisfied up to ${\cal O}(\tilde\alpha^2)$.

Off-diagonal entries of $(m_{\tilde L}^2)_{ij}^{(SCKM)}$
 and   $(m_{\tilde R}^2)_{ij}^{(SCKM)}$
are constrained by the FCNC experiments~\cite{Gabbiani:1996hi}.
Our model predicts
\begin{eqnarray}
(\Delta_{LL})_{12} \equiv 
\frac{(m_{\tilde L}^2)_{12}^{(SCKM)}}{(m_{\tilde L}^2)_{11}}= {\cal
  O}({\tilde \alpha}^2), 
\qquad 
(\Delta_{RR})_{12} \equiv \frac{(m_{\tilde R}^2)_{12}^{(SCKM)}}
{(m_{\tilde R}^2)_{11}}= {\cal O}({\tilde \alpha}^2),
\end{eqnarray} 
where we take $m_L^2=m_{R(1)}^2=m_{3/2}^2$.
The $\mu \rightarrow e \gamma$ experiment~\cite{Adam:2009ci} constrains 
these values as $(\Delta_{LL})^{\rm exp}_{12}$, $(\Delta_{RR})^{\rm exp}_{12} 
\leq {\cal O}(10^{-3})$~\cite{Gabbiani:1996hi}, when $m_{\tilde L}\simeq
 m_{\tilde R} \simeq 100$ GeV.
On the other hand, the parameter space in the previous section 
corresponds to $\tilde \alpha \simeq 10^{-2}$ and 
gives  $(\Delta_{LL})_{12}$, $(\Delta_{RR})_{12}\leq {\cal  O}(10^{-4})$.
Thus, our parameter region is favored from 
the viewpoint of the  FCNC constraint.

\subsection{A-term of slepton sector}

Here, let us study scalar trilinear couplings, i.e. 
the so called A-terms.
The A-terms among left-handed and right-handed sleptons 
and Higgs scalar fields are obtained in the gravity mediation 
as~\cite{Kaplunovsky:1993rd}
\begin{equation}
h_{IJ} {L}_J {R}_I H_K =  \sum_{K={\bar 5},\ 45}
h^{(Y)}_{IJK}{L}_J {R}_I H_K  + h^{(K)}_{IJK}{L}_J {R}_I H_K ,
\label{eq:A-term}
\end{equation}
where 
\begin{eqnarray}
h^{(Y)}_{IJK} &=& F^{\Phi_k} \langle \partial_{\Phi_k} \tilde{y}_{IJK}
\rangle ,  
\nonumber \\
h^{(K)}_{IJK}{L}_J {R}_I H_K &=& - 
\langle \tilde{y}_{LJK} \rangle {L}_J {R}_I H_K F^{\Phi_k} K^{L\bar{L}}
\partial_{\Phi_k} K_{\bar{L}I}  \\
& &  -
\langle \tilde{y}_{IMK} \rangle {L}_J {R}_I H_d F^{\Phi_k} K^{M\bar{M}} 
\partial_{\Phi_k} K_{\bar{M}J}  \nonumber  \\ 
& & -   \langle \tilde{y}_{IJK} \rangle {L}_J {R}_I H_K F^{\Phi_k} K^{H_d}
\partial_{\Phi_k} K_{H_K}, \nonumber  
\label{eq:A-term-2}
\end{eqnarray}
and $K_{H_K}$ denotes the K\"ahler metric of $H_K$.
In addition, $\tilde{y}_{IJK}$ denotes effective Yukawa couplings, 
and it corresponds to 
\begin{eqnarray}
\tilde{y}_{IJK}
 = 
-3y_1
\begin{pmatrix}0 & \alpha _9/\sqrt 2 & -\alpha _{10}/\sqrt 2 \\ 
           -2\alpha _8/\sqrt 6  & \alpha _9/\sqrt 6 & \alpha _{10}/\sqrt 6  \\
                 0  & 0 & 0   \\
 \end{pmatrix} 
+y_2
\begin{pmatrix} 0 & 0 & 0 \\ 
               0 & 0 & 0 \\
                 \alpha _{11} & \alpha _{12} & \alpha _{13} \\
 \end{pmatrix}.
\label{ME}
 \end{eqnarray}
Since the vacuum alignment indicates 
$\alpha _{10}=\alpha _{12}=\alpha _{13}=\alpha _{14}=0$, we get 
\begin{eqnarray}
\tilde{y}_{IJK}
 = 
-3y_1
\begin{pmatrix}0 & \alpha _9/\sqrt 2 & 0 \\ 
           0 & \alpha _9/\sqrt 6 & 0 \\
                 0  & 0 & 0   \\
 \end{pmatrix} 
+y_2
\begin{pmatrix} 0 & 0 & 0 \\ 
               0 & 0 & 0 \\
                 0 & 0 & \alpha _{13} \\
 \end{pmatrix}, 
\label{ME}
 \end{eqnarray}
then 
\begin{eqnarray}
h^{(Y)}_{IJK}
 = 
-\frac{3y_1}{\Lambda }
\begin{pmatrix}0 & \tilde F^{\alpha _9}/\sqrt 2 & 0 \\ 
           0 & \tilde F^{\alpha _9}/\sqrt 6 & 0  \\
                 0  & 0 & 0   \\
 \end{pmatrix}
+\frac{y_2}{\Lambda }
\begin{pmatrix} 0 & 0 & 0 \\ 
               0 & 0 & 0 \\
                0 & 0 & \tilde F^{\alpha _{13}} \\
 \end{pmatrix},
\label{ME}
\end{eqnarray}
where $\tilde F^{\alpha_i}=F^{\alpha_i}/\alpha_i$ and 
$\tilde F^{\alpha_i}/\Lambda = {\cal O}(m_{3/2})$.

Next, we estimate $h^{(K)}_{IJK}$.
When we neglect correction terms and use the lowest level 
of K\"ahler potential, we obtain 
\begin{equation}
h^{(K)}_{IJK} = \tilde y_{IJK} (A^R_I+A^L_J),
\end{equation}
where  
 we estimate $A^L_1=A^L_2=A^L_3 
=F^{\tilde\alpha_i}/(\alpha_i\Lambda) \simeq \mathcal{O}(m_{3/2})$. 
The magnitudes of $A^R_1= A^R_2$ and $A^R_3$ are also $\mathcal{O}(m_{3/2})$.

Furthermore, we should take into account next-to-leading terms of 
the K\"ahler potential including $\chi_i$.
These correction terms appear all entries so that their  magnitudes
 are suppressed in ${\cal O}(\tilde\alpha)$
compared with the leading term.  Then, we obtain 
\begin{equation}
 (m_{LR}^2)_{ij} \simeq m_{3/2}
\begin{pmatrix}
\mathcal{O} \left (\tilde \alpha ^2v_d\right ) 
& \frac{\sqrt{3}m_\mu}{2} & \mathcal{O} \left (\tilde \alpha ^2v_d\right ) \\ 
\mathcal{O} \left (\tilde \alpha ^2v_d\right ) 
& \frac{m_\mu}{2} & \mathcal{O} \left (\tilde \alpha ^2v_d\right ) \\
\mathcal{O} \left (\tilde \alpha ^2v_d\right ) & \mathcal{O} \left (\tilde \alpha ^2v_d\right ) & \mathcal{O}(m_\tau) 
\end{pmatrix}.
\end{equation}

Moving to the super-CKM basis, 
we have
\begin{align}
 (m_{LR}^2)^{SCKM}_{ij} =U_E^\dagger  (m_{LR}^2)_{ij} V_E
&\simeq m_{3/2}
\begin{pmatrix}
\mathcal{O} \left (\tilde \alpha ^2v_d\right ) & \mathcal{O} \left (\tilde \alpha ^2v_d\right ) & \mathcal{O} \left (\tilde \alpha ^2v_d\right ) \\ 
\mathcal{O} \left (\tilde \alpha ^2v_d\right ) 
& \mathcal{O}(m_\mu) & \mathcal{O} \left(\tilde \alpha ^2v_d\right ) \\
\mathcal{O} \left (\tilde \alpha ^2v_d\right ) & \mathcal{O} \left (\tilde \alpha ^2v_d\right ) & \mathcal{O}  (m_\tau)
\end{pmatrix} \nonumber \\
&\simeq m_{3/2}
\begin{pmatrix}
\mathcal{O} \left (m_e\right ) & \mathcal{O} \left (m_e\right ) & 
\mathcal{O} \left (m_e\right ) \\ 
\mathcal{O} \left (m_e\right ) & \mathcal{O}(m_\mu) & 
\mathcal{O} \left ( m_e\right ) \\
\mathcal{O} \left (m_e\right ) & \mathcal{O} \left (m_e\right ) 
& \mathcal{O}  (m_\tau)
\end{pmatrix}.
\end{align}
The FCNC measure 
\begin{equation}
(\Delta_{LR})_{12}\equiv 
\frac{\left (\tilde m_{LR}^2\right )_{12}}{m_{3/2}^2} = 
\frac{{\cal O} (m_e)}{m_{3/2}},
\end{equation} 
is predicted to be of order 
 $5\times 10^{-6}$  for  $m_{3/2} = 100$ GeV.

This ratio is marginal compared with  the bound  ${\cal O}(10^{-6})$  by 
the $\mu \rightarrow e \gamma$ experiments
when the slepton mass is $100$ GeV.
Therefore, we expect  that  the  $\mu \rightarrow e\gamma $
process will observed in the near future if slepton mass is $100$ GeV.

\subsection{FCNC from neutrino sector}
Next, we consider the contribution on FCNC from the neutrino sector.
The typical process of the FCNC is 
 the $\mu \rightarrow e\gamma $~\cite{Adam:2009ci}. 
The key quantity is the magnitude of  $(Y_D^\dagger Y_D)_{12}$,
which is given in  the following matrix:
\begin{align}
Y_D^\dagger Y_D=\frac{\alpha _5^2}{3}
\begin{pmatrix}
3{y_2^D}^2+2{y_1^D}^2\lambda ^2 & 3{y_2^D}^2-{y_1^D}^2\lambda ^2 & 3{y_2^D}^2-{y_1^D}^2\lambda ^2 \\
3{y_2^D}^2-{y_1^D}^2\lambda ^2 & 3{y_2^D}^2+2{y_1^D}^2\lambda ^2 & 3{y_2^D}^2-{y_1^D}^2\lambda ^2 \\
3{y_2^D}^2-{y_1^D}^2\lambda ^2 & 3{y_2^D}^2-{y_1^D}^2\lambda ^2 & 3{y_2^D}^2+2{y_1^D}^2\lambda ^2
\end{pmatrix}
\simeq {y_2^D}^2
\begin{pmatrix}
\alpha _5^2 & \alpha _5^2 & \alpha _5^2 \\
\alpha _5^2 & \alpha _5^2 & \alpha _5^2 \\
\alpha _5^2 & \alpha _5^2 & \alpha _5^2 
\end{pmatrix}.
\end{align}
The FCNC measure on $\mu \rightarrow e\gamma $ is calculated  as follows~\cite{Borzumati:1986qx,Hisano:1995nq,Hisano:1995cp}:
\begin{equation}
(\Delta_{LL})_{12}\equiv
 \frac{\left (\Delta m\right )_{12}^2}{M_\text{SUSY}^2}=
\frac{6m_0^2}{16\pi ^2M_\text{SUSY}^2}(Y_D^\dagger Y_D)_{12}
\ln \frac{\Lambda }{M}
\simeq \frac{3}{8\pi^2}{y_2^D}^2\alpha _5^2 \ln \frac{\Lambda }{M}
\simeq 6\times 10^{-5}\ ,
\end{equation}
where we put 
$m_0=M_\text{SUSY}$, $\alpha _5=10^{-2}$, 
$\Lambda =10^{18}$ GeV,  $M=10^{12}$ GeV.
 It is concluded that the contribution 
 on  $\mu \rightarrow e\gamma $ from the neutrino sector is 
much smaller than the experimental bound  
$(\Delta_{LL})^{\rm exp}_{12} \leq {\cal O}(10^{-3})$~\cite{Gabbiani:1996hi}.

%and consider the next-to-leading order of $Y_D^\dagger Y_D$, we get 
%\begin{scriptsize}
%\begin{align}
%&Y_D^\dagger Y_D=\frac{1}{6}\times  \nonumber \\
%&\hspace{-7mm}\begin{pmatrix}
%2\alpha _7^2\left (3{y_2^D}^2+2{y_1^D}^2\lambda ^2\right )+6{y_\Delta ^D}^2\lambda ^2\alpha _{11}^2\alpha _{15}^2  
%& 2\alpha _7^2\left (3{y_2^D}^2-{y_1^D}^2\lambda ^2\right )+3\sqrt 2y_1^Dy_\Delta ^D\lambda ^2\alpha _7\alpha _{11}\alpha _{15} 
%& 2\alpha _7^2\left (3{y_2^D}^2-{y_1^D}^2\lambda ^2\right )-3\sqrt 2y_1^Dy_\Delta ^D\lambda ^2\alpha _7\alpha _{11}\alpha _{15} \\
%2\alpha _7^2\left (3{y_2^D}^2-{y_1^D}^2\lambda ^2\right )+3\sqrt 2y_1^Dy_\Delta ^D\lambda ^2\alpha _7\alpha _{11}\alpha _{15} 
%& 2\alpha _7^2\left (3{y_2^D}^2+2{y_1^D}^2\lambda ^2\right ) & 2\alpha _7^2\left (3{y_2^D}^2-{y_1^D}^2\lambda ^2\right ) \\
%2\alpha _7^2\left (3{y_2^D}^2-{y_1^D}^2\lambda ^2\right )-3\sqrt 2y_1^Dy_\Delta ^D\lambda ^2\alpha _7\alpha _{11}\alpha _{15} 
%& 2\alpha _7^2\left (3{y_2^D}^2-{y_1^D}^2\lambda ^2\right ) & 2\alpha _7^2\left (3{y_2^D}^2+2{y_1^D}^2\lambda ^2\right )
%\end{pmatrix}.
%\end{align}
%\end{scriptsize}

\section{Summary}

We have presented  a flavor model with the  $S_4$ symmetry  to unify
  quarks and leptons in the framework of the $SU(5)$ SUSY GUT.
 Three generations of $\overline 5$-plets in $SU(5)$ are assigned to ${\bf 3}$ of 
$S_4$ while the  first and second generations of 
$10$-plets  in  $SU(5)$  are assigned to ${\bf 2}$ of $S_4$,
and the third generation of $10$-plet is assigned to ${\bf 1}$ of $S_4$.
These  assignments of $S_4$ for $\overline 5$ and $10$ 
lead to the  completely different structure 
of  quark and lepton mass matrices.
Right-handed neutrinos, which are $SU(5)$ gauge singlets, 
are also assigned to ${\bf 2}$ for the first and second generations
and ${\bf 1}'$ for  the third generation. 
These  assignments  realize the tri-bimaximal mixing
of neutrino flavors.
The vacuum alignment of scalars is also required to realize the tri-bimaximal 
mixing of neutrino flavors.
Our model predicts the quark  mixing  as well as the tri-bimaximal
mixing of leptons. Especially, the Cabibbo angle is
predicted to be around    $15^{\circ}$.
Our model is consistent with   observed CKM mixing angles
and $CP$ violation
 as well as the non-vanishing $U_{e3}$ of the neutrino flavor mixing.
The deviation from  $15^{\circ}$ in  $|V_{us}^0|$ is given by
 ${\cal O}({m_d/m_s})$. Therefore, we can adjust one  parameter
at the next-to-leading order   to reproduce the observed Cabibbo angle.
The non-vanishing $U_{e3}$ of the neutrino flavor mixing
is also predicted to be $\sim 0.02$.

We have also studied SUSY breaking terms.
In our model, three families of left-handed slepton masses 
are degenerate and 
two right-handed sleptons   are degenerate.
Even although we take into account corrections due to the 
flavor symmetry breaking, our model leads to 
marginal values of FCNC's compared with the present experimental bounds.
Therefore, we expect the observation of  the  $\mu \rightarrow e\gamma $
process in the near future.

%%%%% acknowledgement %%%%%
\vspace{1cm}
\noindent
{\bf Acknowledgement}

We thank T. Kobayashi for useful discussion of  soft SUSY breaking.
H.I. is supported by Grand-in-Aid for Scientific Research,
No.21.5817 from the Japan Society of Promotion of Science.
The work of M.T. is  supported by the
Grant-in-Aid for Science Research, No. 21340055,
from the Ministry of Education, Culture,
Sports, Science and Technology of Japan.

\newpage
\noindent
{\LARGE \bf Appendix}
\appendix

\section{Multiplication rule of $S_4$}

The $S_4$ group has 24 distinct elements and irreducible representations 
${\bf 1},~{\bf 1}',~{\bf 2},~{\bf 3}$, and ${\bf 3}'$.
The multiplication rule depends on the basis.
One can see its  basis dependence in our  review~\cite{Ishimori:2010au}.
In this appendix, we present  the multiplication rule, 
which is used in this paper:
\begin{align}
\begin{pmatrix}
a_1 \\
a_2
\end{pmatrix}_{\bf 2} \otimes  \begin{pmatrix}
                                      b_1 \\
                                      b_2
                                  \end{pmatrix}_{\bf 2}
 &= (a_1b_1+a_2b_2)_{{\bf 1}}  \oplus (-a_1b_2+a_2b_1)_{{\bf 1}'} 
  \oplus  \begin{pmatrix}
             a_1b_2+a_2b_1 \\
             a_1b_1-a_2b_2
         \end{pmatrix}_{{\bf 2}\ ,} \\
\begin{pmatrix}
a_1 \\
a_2
\end{pmatrix}_{\bf 2} \otimes  \begin{pmatrix}
                                      b_1 \\
                                      b_2 \\
                                      b_3
                                  \end{pmatrix}_{{\bf 3}}
 &= \begin{pmatrix}
          a_2b_1 \\
          -\frac{1}{2}(\sqrt 3a_1b_2+a_2b_2) \\
          \frac{1}{2}(\sqrt 3a_1b_3-a_2b_3)
      \end{pmatrix}_{{\bf 3}} \oplus \begin{pmatrix}
                                        a_1b_1 \\
                                        \frac{1}{2}(\sqrt 3a_2b_2-a_1b_2) 
\\
                                        -\frac{1}{2}(\sqrt 3a_2b_3+a_1b_3)
                                   \end{pmatrix}_{{\bf 3}'\ ,} \\
\begin{pmatrix}
a_1 \\
a_2
\end{pmatrix}_{\bf 2} \otimes  \begin{pmatrix}
                                      b_1 \\
                                      b_2 \\
                                      b_3
                                  \end{pmatrix}_{{\bf 3}'}
&= \begin{pmatrix}
         a_1b_1 \\
         \frac{1}{2}(\sqrt 3a_2b_2-a_1b_2) \\
         -\frac{1}{2}(\sqrt 3a_2b_3+a_1b_3)
     \end{pmatrix}_{{\bf 3}} \oplus
      \begin{pmatrix}
                                      a_2b_1 \\
                                      -\frac{1}{2}(\sqrt 3a_1b_2+a_2b_2) \\
                                      \frac{1}{2}(\sqrt 3a_1b_3-a_2b_3)
                                  \end{pmatrix}_{{\bf 3}'\ ,} \\
\begin{pmatrix}
a_1 \\
a_2 \\
a_3
\end{pmatrix}_{{\bf 3}} \otimes  \begin{pmatrix}
                                      b_1 \\
                                      b_2 \\
                                      b_3
                                  \end{pmatrix}_{{\bf 3}}
 &= (a_1b_1+a_2b_2+a_3b_3)_{{\bf 1}} 
  \oplus \begin{pmatrix}
             \frac{1}{\sqrt 2}(a_2b_2-a_3b_3) \\                                            
             \frac{1}{\sqrt 6}(-2a_1b_1+a_2b_2+a_3b_3)
         \end{pmatrix}_{\bf 2} \nonumber \\
 &\ \oplus \begin{pmatrix}
            a_2b_3+a_3b_2 \\
            a_1b_3+a_3b_1 \\
            a_1b_2+a_2b_1
         \end{pmatrix}_{{\bf 3}} \oplus \begin{pmatrix}
                                          a_3b_2-a_2b_3 \\
                                          a_1b_3-a_3b_1 \\
                                          a_2b_1-a_1b_2
                                       \end{pmatrix}_{{\bf 3}'\ ,} \\
\begin{pmatrix}
a_1 \\
a_2 \\
a_3
\end{pmatrix}_{{\bf 3}'} \otimes  \begin{pmatrix}
                                      b_1 \\
                                      b_2 \\
                                      b_3
                                  \end{pmatrix}_{{\bf 3}'}
 &= (a_1b_1+a_2b_2+a_3b_3)_{{\bf 1}}
  \oplus \begin{pmatrix}
             \frac{1}{\sqrt 2}(a_2b_2-a_3b_3) \\                                            
             \frac{1}{\sqrt 6}(-2a_1b_1+a_2b_2+a_3b_3)
         \end{pmatrix}_{\bf 2} \nonumber \\
 &\ \oplus \begin{pmatrix}
            a_2b_3+a_3b_2 \\
            a_1b_3+a_3b_1 \\
            a_1b_2+a_2b_1
         \end{pmatrix}_{{\bf 3}} \oplus \begin{pmatrix}
                                          a_3b_2-a_2b_3 \\
                                          a_1b_3-a_3b_1 \\
                                          a_2b_1-a_1b_2
                                       \end{pmatrix}_{{\bf 3}'\ ,} \\
\begin{pmatrix}
a_1 \\
a_2 \\
a_3
\end{pmatrix}_{{\bf 3}} \otimes  \begin{pmatrix}
                                      b_1 \\
                                      b_2 \\
                                      b_3
                                  \end{pmatrix}_{{\bf 3}'}
 &= (a_1b_1+a_2b_2+a_3b_3)_{{\bf 1}'}  
 \oplus \begin{pmatrix}
             \frac{1}{\sqrt 6}(2a_1b_1-a_2b_2-a_3b_3) \\
             \frac{1}{\sqrt 2}(a_2b_2-a_3b_3)
         \end{pmatrix}_{\bf 2} \nonumber \\
 &\ \oplus \begin{pmatrix}
            a_3b_2-a_2b_3 \\
            a_1b_3-a_3b_1 \\
            a_2b_1-a_1b_2
         \end{pmatrix}_{{\bf 3}} \oplus \begin{pmatrix}
                                          a_2b_3+a_3b_2 \\
                                          a_1b_3+a_3b_1 \\
                                          a_1b_2+a_2b_1
                                       \end{pmatrix}_{{\bf 3}'\ .}
\end{align}

 More details are shown in the review~\cite{Ishimori:2010au}.

\section{Determination of $\ell $, $m$, and $n$}

%The up-type quark masses in Eq.~(\ref{uptypequarkmass}) give
%\begin{equation}
%\alpha _2=\sqrt{\frac{m_um_c}{{y_1^u}^2\lambda ^{4\ell}v_u^2}}\ .
%\end{equation}
%If we take $y_1^u=1$, $\lambda =0.1$, $m_u=1.04$ MeV and $m_c=302$ MeV, 
%we get
%\begin{equation}
%\alpha _2\sim 1.074\times 10^{2\ell -4}\ .
%\end{equation}
%Suppose that magnitudes of all $\alpha_i$ are same order 
%$10^{-2}$ as seen in subsection 2.4.
%In order to take  $\alpha _2\sim \mathcal{O}(10^{-2})$, 
% $\ell $ is fixed as follows;
%\begin{align}
%2\ell -4&=-2 \nonumber \\
%\ell &=1\ .
%\end{align}

The charged lepton masses in Eq.~(\ref{chargemass}) give
\begin{equation}
\alpha _9=\frac{m_\mu }{\sqrt 6|\bar y_1|\lambda ^\ell v_d}\ .
\end{equation}
Therefore, if we take $|\bar y_1|=1$, 
$\lambda =0.1$, and $m_\mu =6.86\times 10^{-2}$ GeV, we get  
\begin{equation}
\alpha _9\sim 5.1\times 10^{\ell -4}\ .
\end{equation}
Suppose  that magnitudes of all $\alpha_i$ are same order $10^{-2}$ as seen in 
Eq.~(\ref{valuealpha}).
 So we take $\ell=1$, which gives  $\alpha _9\sim 5.1\times 10^{-3}$.

We take the right-handed neutrino mass $M$ as follows:
\begin{equation}
M= \mathcal{O}(10^{12})\ \text{GeV}\ .
\end{equation}
As seen in Eqs.~(\ref{neutrinomassparameter}) and (\ref{mass123}), 
 parameters $a$ and $c$ should be comparable. Therefore, we have
\begin{equation}
y_2^N\lambda ^{-n}\alpha _4\Lambda \sim M=\mathcal{O}(10^{12})\  \text{GeV}\ .
\end{equation}
Again suppose that magnitudes of all $\alpha_i$ are same order $10^{-2}$, 
we get following equation as
\begin{equation}
y_2^N\Lambda \sim \mathcal{O}(10^{14-n})\ .
\label{cut-off}
\end{equation}
where $\lambda =0.1$.  
On the other hand,  $m$ and $n$ satisfy the condition:
\begin{equation}
0<m<n\leq 2m\ .
\label{m-n-condition}
\end{equation}
Therefore, we have 
 the smallest value of $n$ which satisfies  Eq.~(\ref{m-n-condition}) as 
\begin{equation}
m=1,\quad n=2\ .
\end{equation}

\section{Next-to-leading terms of the scalar potential}

 We consider the next-to-leading terms of the scalar potential. 
In the next-to-leading order, 
the $SU(5)\times S_4\times Z_4\times U(1)_{FN}\times U(1)_R$ 
invariant operators which couple  to driving fields are given as follows;
\begin{itemize}
\item Coupled with $\chi _1^0$:
\begin{align}
&\left (\chi _1,\chi _2\right )\otimes \left (\chi _1,\chi _2\right )\otimes \chi _1^0\otimes \Theta ^n/\bar \Lambda ^n, \nonumber \\
&\ \chi _{14}\otimes \chi _{14}\otimes \chi _1^0\otimes \Theta ^n/\bar \Lambda ^n, \nonumber \\
&\left (\chi _5,\chi _6,\chi _7\right )\otimes \left (\chi _5,\chi _6,\chi _7\right )\otimes \chi _1^0\otimes \Theta ^{2\ell +n}/\bar \Lambda ^{2\ell +n}, \nonumber \\
&\left (\chi _{11},\chi _{12},\chi _{13}\right )\otimes \left (\chi _{11},\chi _{12},\chi _{13}\right )\otimes \chi _1^0
\otimes \Theta ^{2\ell +n}/\bar \Lambda ^{2\ell +n}, \nonumber \\
&\left (\chi _3,\chi _4\right )\otimes \left (\chi _5,\chi _6,\chi _7\right )\otimes \left (\chi _5,\chi _6,\chi _7\right )\otimes \chi _1^0
\otimes \Theta ^{2\ell }/(\Lambda \bar \Lambda ^{2\ell }), \nonumber \\
&\left (\chi _3,\chi _4\right )\otimes \left (\chi _{11},\chi _{12},\chi _{13}\right )\otimes \left (\chi _{11},\chi _{12},\chi _{13}\right )\otimes \chi _1^0
\otimes \Theta ^{2\ell }/(\Lambda \bar \Lambda ^{2\ell }), \nonumber \\
&\left (\chi _5,\chi _6,\chi _7\right )\otimes \left (\chi _8,\chi _9,\chi _{10}\right )\otimes \left (\chi _{11},\chi _{12},\chi _{13}\right )\otimes \chi _1^0
\otimes \Theta ^{2\ell +n}/(\Lambda \bar \Lambda ^{2\ell +n}), \nonumber \\
&\left (\chi _8,\chi _9,\chi _{10}\right )\otimes \left (\chi _8,\chi _9,\chi _{10}\right )\otimes \left (\chi _8,\chi _9,\chi _{10}\right )\otimes \chi _1^0
\otimes \Theta ^{2\ell +n}/(\Lambda \bar \Lambda ^{2\ell +n}).
\label{next-to-leading}
\end{align}
\item Coupled with $\chi _2^0$:
\begin{align}
&\left (\chi _5,\chi _6,\chi _7\right )\otimes \left (\chi _5,\chi _6,\chi _7\right )\otimes \left (\chi _{11},\chi _{12},\chi _{13}\right )
\otimes \chi _2^0/\Lambda , \nonumber \\
&\left (\chi _5,\chi _6,\chi _7\right )\otimes \left (\chi _8,\chi _9,\chi _{10}\right )\otimes \left (\chi _8,\chi _9,\chi _{10}\right )
\otimes \chi _2^0/\Lambda , \nonumber \\
&\left (\chi _{11},\chi _{12},\chi _{13}\right )\otimes \left (\chi _{11},\chi _{12},\chi _{13}\right )\otimes \left (\chi _{11},\chi _{12},\chi _{13}\right )
\otimes \chi _2^0/\Lambda .
\end{align}
\item Coupled with $\chi _3^0$:
\begin{align}
&\left (\chi _5,\chi _6,\chi _7\right )\otimes \left (\chi _5,\chi _6,\chi _7\right )\otimes \chi _3^0
\otimes \Theta ^{2\ell }/\bar \Lambda ^{2\ell }, \nonumber \\
&\left (\chi _{11},\chi _{12},\chi _{13}\right )\otimes \left (\chi _{11},\chi _{12},\chi _{13}\right )\otimes \chi _3^0
\otimes \Theta ^{2\ell }/\bar \Lambda ^{2\ell }, \nonumber \\
&\left (\chi _3,\chi _4\right )\otimes \left (\chi _5,\chi _6,\chi _7\right )\otimes \left (\chi _5,\chi _6,\chi _7\right )\otimes \chi _3^0
\otimes \Theta ^{2\ell -n}/(\Lambda \bar \Lambda ^{2\ell -n}), \nonumber \\
&\left (\chi _3,\chi _4\right )\otimes \left (\chi _{11},\chi _{12},\chi _{13}\right )\otimes \left (\chi _{11},\chi _{12},\chi _{13}\right )\otimes \chi _3^0
\otimes \Theta ^{2\ell -n}/(\Lambda \bar \Lambda ^{2\ell -n}), \nonumber \\
&\left (\chi _5,\chi _6,\chi _7\right )\otimes \left (\chi _8,\chi _9,\chi _{10}\right )\otimes \left (\chi _{11},\chi _{12},\chi _{13}\right )\otimes \chi _3^0
\otimes \Theta ^{2\ell }/(\Lambda \bar \Lambda ^{2\ell }) ,\nonumber \\
&\left (\chi _8,\chi _9,\chi _{10}\right )\otimes \left (\chi _8,\chi _9,\chi _{10}\right )\otimes \left (\chi _8,\chi _9,\chi _{10}\right )\otimes \chi _3^0
\otimes \Theta ^{2\ell }/(\Lambda \bar \Lambda ^{2\ell }).
\end{align}
\item Coupled with $(\chi _4^0,\chi _5^0)$:
\begin{align}
&\left (\chi _8,\chi _9,\chi _{10}\right )\otimes \left (\chi _{11},\chi _{12},\chi _{13}\right )\otimes \left (\chi _{15},\chi _{16},\chi _{17}\right )
\otimes (\chi _4^0,\chi _5^0)/\Lambda ,\nonumber \\
&\left (\chi _5,\chi _6,\chi _7\right )\otimes \left (\chi _5,\chi _6,\chi _7\right )\otimes \left (\chi _{11},\chi _{12},\chi _{13}\right )
\otimes \left (\chi _{15},\chi _{16},\chi _{17}\right )\otimes (\chi _4^0,\chi _5^0)/\Lambda ^2, \nonumber \\
&\left (\chi _5,\chi _6,\chi _7\right )\otimes \left (\chi _8,\chi _9,\chi _{10}\right )\otimes \left (\chi _8,\chi _9,\chi _{10}\right )\otimes 
\otimes \left (\chi _{15},\chi _{16},\chi _{17}\right )\otimes (\chi _4^0,\chi _5^0)/\Lambda ^2 ,\nonumber \\
&\left (\chi _{11},\chi _{12},\chi _{13}\right )\otimes \left (\chi _{11},\chi _{12},\chi _{13}\right )\otimes \left (\chi _{11},\chi _{12},\chi _{13}\right )
\otimes \left (\chi _{15},\chi _{16},\chi _{17}\right )\otimes (\chi _4^0,\chi _5^0)/\Lambda ^2 .
\end{align}
\end{itemize}
As seen in  Eqs.~(\ref{scalar-leading}) and (\ref{next-to-leading}), 
we can write  the superpotential which couples to  $\chi _1^0$ as 
\begin{align}
w_{\chi _1^0} &= \kappa _1\left (\chi _1,\chi _2\right )\otimes \left (\chi _1,\chi _2\right )\otimes \left (\chi _3,\chi _4\right )\otimes \chi _1^0/\Lambda \nonumber \\
&\ + \kappa _2\left (\chi _1,\chi _2\right )\otimes \left (\chi _1,\chi _2\right )\otimes \chi _1^0\otimes \Theta ^n/\bar \Lambda ^n 
 + \kappa _3\chi _{14}\otimes \chi _{14}\otimes \chi _1^0\otimes \Theta ^n/\bar \Lambda ^n \ ,
\end{align}
which is rewritten as 
\begin{align}
w_{\chi _1^0} &= \kappa _1\left [2\chi _1\chi _2\chi _3+\left (\chi _1^2-\chi _2^2\right )\chi _4\right ]\chi _1^0/\Lambda
 + \kappa _2\lambda ^n(\chi _1^2+\chi _2^2)\chi _1^0 + \kappa _3\lambda ^n\chi _{14}^2\chi _1^0 \ .
\end{align}
Then, we obtain
\begin{align}
\kappa _1\left [2\chi _1\chi _2\chi _3+\left (\chi _1^2-\chi _2^2\right )\chi _4\right ]/\Lambda  
+ \kappa _2\lambda ^n(\chi _1^2+\chi _2^2) + \kappa _3\lambda ^n\chi _{14}^2 = 0 \ .
\end{align}
Inserting $\chi _{14}^2=-\frac{\eta _2}{\eta _3}(\chi _1^2+\chi _2^2)$ 
in Eq.~(\ref{vacuum-alignment}) into this equation, we have 
\begin{align}
\left [\left (\kappa _2-\kappa _3\frac{\eta _2}{\eta _3}\right )\lambda ^n+\kappa _1\chi _4/\Lambda \right ]\chi _1^2
+\left [\left (\kappa _2-\kappa _3\frac{\eta _2}{\eta _3}\right )\lambda ^n-\kappa _1\chi _4/\Lambda \right ]\chi _2^2
+2\kappa _1\chi _1\chi _2\chi _3/\Lambda =0\ .
\end{align}
Taking $\chi _3\simeq 0$, we get 
\begin{equation}
\frac{\chi _1^2}{\chi _2^2}=\frac{\kappa _1\alpha _4-\left (\kappa _2-\kappa _3\frac{\eta _2}{\eta _3}\right )\lambda ^n}
{\kappa _1\alpha _4+\left (\kappa _2-\kappa _3\frac{\eta _2}{\eta _3}\right )\lambda ^n}\ .
\end{equation}
As far as  $\kappa_1\alpha_4$ is much larger than 
$(\kappa_2-\kappa _3\frac{\eta _2}{\eta _3})\lambda^n$,
 the vacuum alignment $\chi_1=\chi_2$ is guaranteed approximately.
Therefore,  magnitude of deviation of the vacuum alignment 
is parameter dependent.
 Calculating other next-to-leading operators, 
we  can find easily that other vacuum alignment is deviated at most of order 
$\tilde \alpha\sim 0.01$.

\end{document}